\documentclass[sigconf]{acmart}

\usepackage{booktabs} 
\usepackage{array}
\usepackage{colortbl}

\usepackage[normalem]{ulem}
\AtBeginDocument{%
  }

\begin{document}

\title[Robots that Evolve with Us]{Robots that Evolve with Us: Modular Co-Design for Personalization, Adaptability, and Sustainability}

\copyrightyear{2026}
\acmYear{2026}
\setcopyright{cc}
\setcctype{by-nc-nd}
\acmConference[CHI '26]{Proceedings of the 2026 CHI Conference on Human Factors in Computing Systems}{April 13--17, 2026}{Barcelona, Spain}
\acmBooktitle{Proceedings of the 2026 CHI Conference on Human Factors in Computing Systems (CHI '26), April 13--17, 2026, Barcelona, Spain}
\acmPrice{}
\acmDOI{10.1145/3772318.3790991}
\acmISBN{979-8-4007-2278-3/2026/04}

\author{Lingyun Chen}
\affiliation{
  \institution{Luddy School of Informatics, Computing, and Engineering, Indiana University Bloomington}
  \city{Bloomington}
  \state{IN}
  \country{USA}
}
  \email{lch2@iu.edu}

\author{Qing Xiao}
\affiliation{
  \institution{Human-Computer Interaction Institute, Carnegie Mellon University}
  \city{Pittsburgh}
  \state{PA}
  \country{USA}
}
  \email{qingx@andrew.cmu.edu}

\author{Zitao Zhang}
\affiliation{
  \institution{Luddy School of Informatics, Computing, and Engineering, Indiana University Bloomington}
  \city{Bloomington}
  \state{IN}
  \country{USA}
}
  \email{zhangzit@iu.edu}

\author{Eli Blevis}
\affiliation{
  \institution{Luddy School of Informatics, Computing, and Engineering, Indiana University Bloomington}
  \city{Bloomington}
  \state{IN}
  \country{USA}
}
  \email{eblevis@iu.edu}

\author{Selma Šabanović}
\affiliation{
  \institution{Luddy School of Informatics, Computing, and Engineering, Indiana University Bloomington}
  \city{Bloomington}
  \state{IN}
  \country{USA}
}
  \email{selmas@iu.edu}

\newcommand{\amendment}[1]{\textcolor{black}{#1}}
\newcommand{\amendmentrmv}[1]{\sout{#1}}

\begin{abstract}
Many current robot designs prioritize efficiency and one-size-fits-all solutions, oftentimes overlooking personalization, adaptability, and sustainability. To explore alternatives, we conducted two co-design workshops with 23 participants, who engaged with a modular robot co-design framework. Using components we provided as building blocks, participants combined, removed, and invented modules to envision how modular robots could accompany them from childhood through adulthood and into older adulthood. The participants’ designs illustrate how modularity (a) enables personalization through open-ended configuration, (b) adaptability across shifting life-stage needs, and (c) sustainability through repair, reuse, and continuity. We therefore derive design principles that establish modularity as a foundation for lifespan-oriented human–robot interaction. This work reframes modular robotics as a flexible and expressive co-design approach, supporting robots that evolve with people, rather than static products optimized for single moments or contexts of use.

\end{abstract}




\maketitle

\section{Introduction}

Robots are increasingly envisioned not as short-term tools but as long-term companions that support people throughout their lives \cite{leite2013social, kaplan2005everyday,chen20253r}. From childhood education to workplace productivity and eldercare \cite{chen20253r, bemelmans2012socially, kory2019exploring, westlund2018measuring, yi2025building}, socially assistive robots hold the promise of providing functional help \cite{deutsch2019home, klamer2010acceptance}, emotional connection \cite{vsabanovic2013paro}, and continuity of care across the human lifespan \cite{irfan2023lifelong, kidd2008robots}. Realizing this promise, however, requires moving beyond one-size-fits-all systems. Because people’s needs, values, and contexts shift over time 
\cite{carstensen1999taking, fung2001age}, robots should remain not only useful at one life stage but also adaptable over decades \cite{de2015sharing}. This raises a central challenge for Human–Robot Interaction (HRI): how can robots remain relevant, meaningful, and sustainable as peoples’ lives advance?

Prior research has begun to address aspects of this challenge, but typically in isolation. Studies on \textit{personalization} emphasize tailoring robots to individual preferences \cite{Bahar2019Personalization, lee2012personalization}, yet these efforts often assume short-term customization rather than long-term relevance. Work on \textit{adaptability} has explored technical self-reconfiguration, such as modular robots capable of rearranging their forms \cite{yim2000polybot}, but with little attention to how adaptability is perceived or enacted by people. Research on \textit{sustainability} has focused on ecological efficiency or robots promoting pro-environmental behavior \cite{blevis2007sustainable, scheutz2021envirobots}, while overlooking how sustained human–robot relationships might themselves prevent obsolescence. What is missing is a holistic design approach in HRI that treats personalization, adaptability, and sustainability not as separate concerns but as interconnected dimensions of designing for lifespan-oriented interaction.

A very recent speculative design framework has begun to bridge these fragmented strands by highlighting the interdependence of \textit{Sustainability, Adaptability, and Modularity (SAM)} \cite{chen2025sustainable}. In this account, sustainability is positioned as the overarching aim, extending product lifespans and reducing waste, while modularity functions as the mechanism that enables repair, upgrading, and personalization. Adaptability, in turn, reflects a robot’s capacity to evolve alongside users’ shifting social, emotional, and functional needs. Crucially, these dimensions are not separate but mutually reinforcing: modularity enables adaptation, adaptation supports long-term use, and sustained use advances sustainability.

While SAM \cite{chen2025sustainable} remains a speculative, system-level design theory focused on ideal robot properties, we build on this work by naming a second complementary framework. Herein, we introduce the Personalization, Adaptability, Sustainability (PAS) framework as a complementary, human-centered model derived from empirical co-design, specifically capturing the lived experience of lifespan HRI. PAS reframes its own three dimensions from within participants’ imagined, ongoing relationship with a single modular robot: personalization captures how people use modules to inscribe their own roles, preferences, and values into the robot; adaptability describes how they reconfigure its form and functions as life events and responsibilities shift; and sustainability names the material and emotional continuity that emerges when modules are repaired, reused, and carried forward over time. In this view, modularity is no longer a separate pillar but the socio-technical practice of assembling, swapping \cite{seo2019modular,zou2022towards,denivsa2023technology}, and recombining parts through which sustainability, adaptability, and personalization are actually enacted in people’s lives.

In this paper, we ask:\textit{ How can modular co-design support personalization, adaptability, and sustainability across the human lifespan in future robot design?} To explore this question, we conducted two co-design workshops with 15 younger adults (ages 21–32) and 8 older adults (65+). Participants were asked to imagine how a single modular robot could accompany them from childhood through adulthood and into older adulthood, reconfiguring through interchangeable modules (heads, arms, torsos, and bases) to meet changing needs. Our findings highlight three mutually reinforcing qualities of modular design. Personalization enabled participants to configure robots that reflected their unique values and circumstances, even within the same life stage. Adaptability allowed these robots to flexibly shift roles, both within a single stage of life (e.g., balancing work and care in adulthood) and across stages (e.g., from playful peer in childhood to supportive companion in older adulthood). Sustainability emerged at the intersection of these qualities: (a) materially, through repair, reuse, and upgrading of modules; and (b) emotionally, through continuity, attachment, and identity over time. These findings show how modularity, which enables user co-design robot practices, transforms robots from disposable gadgets into enduring companions.

This work makes three main contributions to HRI and design community:
\begin{enumerate}
    \item We provide empirical evidence of how users imagine modular robots evolve across their lifespan, addressing a gap in HRI research that has largely studied user groups in isolation. 
    \item We advance a conceptual understanding of HRI durability by introducing an additional human-centered framework, namely Personalization, Adaptability, and Sustainability (PAS) as a complementary framework to SAM. PAS adds focus on the experiential dimensions of long-term human-robot relationships.
    \item We propose design principles for sustainable HCI that leverage modularity as both a technical mechanism and a design philosophy, supporting robots that remain ecologically responsible, emotionally resonant, and meaningful over the long term.
\end{enumerate}

\section{Related Work}
In this section, we review two strands of prior work that inform our study: (1) research on sustainability in Human–Robot Interaction and the emerging role of modular robotics (see \autoref{lr1}), and (2) research on robots across the human lifespan (see \autoref{lr2}). These perspectives highlight both the ecological and experiential dimensions of sustainability, and point to the need for human-centered approaches that envision robots adapting with people over time.  

\subsection{Sustainable HRI and Modular Robots} \label{lr1}

Concerns about sustainability have long shaped Human–Computer Interaction (HCI), where frequent cycles of consumption and technological obsolescence generate significant environmental impacts \cite{cooper2012longer, disalvo2010mapping, hansson2021decade, remy2014addressing, yu2024if}. In 2007, \citeauthor{blevis2007sustainable} introduced the concept of Sustainable Interaction Design (SID), also called Sustainable HCI (SHCI) \cite{hansson2021decade}, calling sustainability to be a central concern of design \cite{blevis2007sustainable}. \citeauthor{blevis2007sustainable}'s framework \cite{blevis2007sustainable} emphasized that interaction design choices have material consequences, from rare metal mining and energy consumption to the accumulation of toxic electronic waste. 

Building on this tradition, Human–Robot Interaction (HRI) has more recently turned toward sustainability \cite{hartmann2021becoming, ye2025game, vsabanovic2010robots}. Much of this work has addressed the environmental costs of rapid technological replacement and disposal \cite{mcgloin2024consulting}. For example, \citeauthor{ye2025game} examined how people’s relationships with robots, as friends, assistants, or pets, influence whether they choose to repair or replace them \cite{ye2025game}. Likewise, \citeauthor{scheutz2021envirobots} proposed “EnviRobots,” embedding psychological strategies such as feedback and prompting into robots to encourage sustainable behaviors like recycling and energy conservation \cite{scheutz2021envirobots}. This research demonstrates that robots can not only perform ecological tasks but also shape human practices in ways that sustain long-term pro-environmental behavior.  

Yet, most current robots are designed as rigid, closed systems \cite{vsabanovic2010robots, koike2024sprout,tang2025robolinker}. They cannot be easily upgraded, adapted, or repaired, leading to cycles of replacement and waste. Modular robotics recently offers a promising alternative \cite{alattas2019evolutionary}. By enabling individual components to be swapped, recombined, or repaired, modular systems can extend the useful life of a robot while supporting personalization and iterative adaptation \cite{alattas2019evolutionary, article}. Technical research has already shown capacities such as self-assembly, self-reconfiguration, self-repair, and self-reproduction \cite{alattas2019evolutionary}. Classic examples include PolyBot, which could reconfigure itself to adapt to different terrains \cite{yim2000polybot}; Roombots, which transform into furniture for reuse in everyday contexts \cite{sproewitz2009roombots, 5508713}; and EMeRGE, an open modular platform that supports continuous experimentation and incremental innovation \cite{moreno2021emerge}. These systems illustrate how modularity underpins resilience and long-term adaptability in robotics. However, sustainability in robotics should not only be understood in terms of material efficiency or waste reduction. A truly sustainable approach also requires attending to how robots remain meaningful, adaptable, and usable for people over time \cite{gajvsek2020sustainable}. Despite technical advances, modular robotics research has largely focused on engineering feasibility, with limited attention to how modularity aligns with human needs, values, and practices across the lifespan. 

To address this gap, we draw on the robot design theory of Sustainability, Adaptability, and Modularity (SAM) \cite{chen2025sustainable}. SAM \cite{chen2025sustainable} conceptualizes Sustainability, Adaptability, and Modularity as three mutually reinforcing pillars of future robot design. Beyond this conceptual framing, our module set also builds on the collection of interchangeable robot parts and a configurable “robot ecology” of heads, arms, bodies, and bases \cite{chen2025sustainable}. However, SAM remains primarily a speculative design framework: it outlines a conceptual constellation but does not provide methods for engaging users or empirically grounding these principles.

By contrast, our work proposes a distinct, human-centered framework of Modular Co-Design for Personalization, Adaptability, and Sustainability (PAS). This framework differs from SAM in two key ways. First, it foregrounds modularity as participatory practice, treating users’ active assembly, disassembly, and reconfiguration as the mechanism through which adaptability and sustainability can be realized. Second, it foregrounds personalization as a human-centered dimension, underscoring the ways in which people modify and adjust robots to fit their changing circumstances across life stages that SAM only hints at rather than fully develops. Personalization, in this sense, is not an additional feature but a theoretical correction: without attending to how robots are tuned to particular people and contexts, the promise of sustainability and adaptability risks remaining abstract and normative.

\subsection{Robots Across the Lifespan} \label{lr2}

HRI studies showed that for robots to be genuinely effective in people’s lives, they should do more than complete short-term tasks \cite{dautenhahn2007socially, kidd2008robots,tang2025robolinker}. They need to adapt to a wide range of situations and provide personalized support that evolves with users over time \cite{umbrico2020holistic,tang2025robolinker,han2024co,han2023robot}. Yet, as \citeauthor{irfan2023lifelong} notes, most HRI research has focused on brief interactions, even though many real-world applications, such as caregiving, education, and companionship, require long-term relationship building between humans and robots \cite{irfan2023lifelong, Bahar2019Personalization}. Existing studies often isolate user groups into narrow categories, such as children or older adults, examining them separately \cite{ahmad2017adaptive, coninx2016towards, yamazaki2023long, ostrowski2022mixed}. While these segmented approaches offer valuable insights into age-specific needs, they tend to overlook how expectations change as people move through different life stages. They also miss the deeper emotional attachments that can develop when robots remain present across longer periods of time. Recent work has further explored how robots might embody long-term temporal change through material processes such as growth, aging, and decay, offering alternative interaction trajectories that unfold over days or months rather than minutes \cite{hu2024designing}. This gap underscores the importance of design approaches that can support sustainable, adaptable, and emotionally resonant engagement throughout the human lifespan.

A life-course perspective addresses this gap by recognizing that people’s relationships with robots evolve continuously as they age, shaped by shifting functional, social, emotional, and cognitive needs \cite{broadbent2009acceptance, kamino2024lifecycle, kamino2025robot, tapus2009use}. Some HCI research suggests that lifelong robotic companions could strengthen emotional attachment, reduce loneliness, and improve continuity of care \cite{kidd2008robots, vsabanovic2010robots}. However, few empirical studies have examined how robot functionalities might transition smoothly across multiple life stages. Current design frameworks rarely account for these dynamic transitions in a holistic way. 

Understanding how robots can adapt fluidly over the lifespan would not only guide the design of long-term interaction strategies, but also highlight the potential for sustained human–robot attachment \cite{kanda2007two, leite2013social}. Such continuity can reinforce sustainability by reducing abandonment, extending product lifespans, and minimizing the need for complete system replacements \cite{kamino2023towards}. In this sense, modular robots hold the potential to achieve what some scholars describe as “heirloom status” \cite{blevis2007sustainable, jung2011deep}, wherein a technology accumulates emotional and symbolic value across generations. By retaining familiar modules and storing shared histories, robots may undergo a process of “ensoulment” \cite{blevis12007ensoulment, nelson2014design, odom2009understanding}, evolving into more than digital devices: they become companions imbued with personal and familial meaning. Much like a grandmother’s wedding ring, a modular robot could one day be cherished and even passed down, carrying both material continuity and emotional resonance.

Our study builds directly on this perspective. By engaging participants in imagining a single modular robot that could accompany them from childhood to adulthood and into older age, we contribute empirical insights into what “lifespan sustainability” might look like in practice. Specifically, we show how modularity can serve as the mechanism that enables adaptation, how adaptation fosters long-term emotional attachment, and how this combination supports sustainability in both ecological and experiential terms. In doing so, our work extends the discussion of robot lifespans from abstract aspiration to concrete human-centered visions, highlighting design opportunities for creating robots that endure, not just as tools, but as lifelong companions.

\section{Methods: Co-Design Workshop about Modular Robots Across Lifespan}
Co-design means designing something together with the people who might one day use it. Instead of asking people only to test a finished product, researchers invite them to take part from the very beginning: sharing ideas, imagining possible futures, and helping shape how a system should look and work. This approach is especially valuable in HRI research \cite{burlando2021robot, gebelli2024co}, wherein robots are often expected to support people over long periods of time and in personal settings such as homes \cite{hofstede2024human, ostrowski2020design, chen20253r}. By making future users active creative partners, co-design helps ensure that robot designs fit people’s real needs and experiences \cite{10.1145/3610977.3634979, garg2022social}. Our study contributes a specific variant of this approach: we use a modular set of physical robot parts as a temporal design probe, inviting participants to repeatedly reconfigure the same robot across imagined life stages. In doing so, modularity itself becomes a research medium for observing how preferences, expectations, and repair practices unfold over a hypothesized lifespan.

In our study, we used co-design to explore how modular robots could grow and change alongside people at different life stages: childhood, adulthood, and older adulthood. To do so, we conducted two in-person workshops in the United States: one with 15 younger adults (ages 21–32) and another with 8 older adults (ages 66–81). This dual-group design allowed us to compare forward-looking imaginations from younger adults with experience-informed reflections from older adults, thereby combining speculative and experiential perspectives on lifespan-oriented modular robotics.  

To guide both workshops, we adopted the robot modules and functions proposed by \citeauthor{chen2025sustainable} in the \textit{2025 ACM Conference on Creativity and Cognition} (see \autoref{fig:modTH}, \autoref{fig:modTA}, \autoref{fig:modTUB}, \autoref{fig:modTLB}). This framework is based on interchangeable building modules, such as heads (H), arms (A), upper bodies (UB), and lower bodies (LB), that can be swapped or combined. We initially kept this structure and four body-like grouping as a starting assumption rather than an end goal---a familiar construction that we expected participants to take up, question, or break. We chose these modules for two reasons. First, they offered a ready-made, functionally diverse toolkit that allowed participants to focus on imagining how roles and relationships change across the lifespan rather than on inventing entirely new mechanical parts from scratch. Second, the familiar “body-like” groupings provided an accessible starting point for participants with no robotics background, while still allowing for playful recombination and distortion.

Importantly, we treated this anthropomorphic scaffold as a starting assumption rather than an end goal. Participants were explicitly encouraged to rearrange, duplicate, or omit parts and to experiment with non-humanoid configurations. As we discuss in our findings, many initially assembled recognizably humanoid robots but then stretched or violated the template—stacking multiple heads, discarding legs, or creating distributed, multi-body systems. This tension between starting from a human-like structure and then breaking it is central to our analysis: it shows how modularity can both reproduce default expectations about robot form and provide a resource for challenging them.

\subsection{Participants}
To investigate how people imagine modular robots evolving across the human lifespan, we conducted two in-person workshops in the United States with participants from distinct age groups. This design allowed us to contrast speculative and experiential perspectives on how one modular robot could accompany a person from childhood, through adulthood, and into older age. All participants followed the same workshop process (see \autoref{set} about detailed workshop set-up), constructing narratives and designs of a single robot that changes its form by swapping modules across life stages.
\begin{table*}[ht]
\caption{Participant Demographics and Experience with Robot}
\centering
%\arrayrulecolor{black}
\small
\renewcommand{\arraystretch}{1.2}
\setlength{\tabcolsep}{4pt}
\begin{tabular}{
  >{\raggedright\arraybackslash}p{0.09\linewidth}  % Participant
  >{\raggedright\arraybackslash}p{0.08\linewidth}  % Gender
  >{\raggedright\arraybackslash}p{0.05\linewidth}  % Age
  >{\raggedright\arraybackslash}p{0.25\textwidth}  % Education
  >{\raggedleft\arraybackslash}p{0.24\textwidth}  % Robot Experience
}
\specialrule{0.12em}{0em}{0em}
\textbf{Participant} & \textbf{Gender} & \textbf{Age} & \textbf{Education} & \textbf{Robot Experience} \\
\specialrule{0.1em}{0em}{0em}
P1 & M & 24 & Masters & Have limited experience \\
P2 & M & 21 & High School & Have used in everyday life \\
P3 & F & 27 & Masters & Have limited experience \\
P4 & F & 27 & Masters & Have limited experience \\
P5 & F & 32 & Masters & Have limited experience \\
P6 & F & 25 & Masters & Have never used \\
P7 & F & 28 & Ph.D. & Have used in everyday life \\
P8 & M & 24 & Bachelors & Have limited experience \\
P9 & F & 24 & Masters & Have limited experience \\
P10 & F & 29 & Ph.D. & Have used in everyday life \\
P11 & F & 27 & Masters & Have limited experience \\
P12 & M & 26 & Masters & Have limited experience \\
P13 & F & 26 & Masters & Have limited experience \\
P14 & F & 26 & Masters & Have limited experience \\
P15 & F & 30 & Masters & Have never used \\
\specialrule{0.1em}{0em}{0em}
P16 & F & 73 & Masters & Have used in everyday life \\
P17 & M & 73 & Bachelors & Have never used \\
P18 & M & 71 & Ph.D. & Have limited experience \\
P19 & F & 68 & Bachelors &  Have limited experience \\
P20 & F & 67 & Masters & Have never used \\
P21 & F & 75 & Bachelors & Have used in everyday life \\
P22 & F & 66 & Masters & Have limited experience\\
P23 & F & 81 & Ph.D. & Have never used \\
\specialrule{0.12em}{0em}{0em}
\end{tabular}
\\
    \footnotesize
    \textit{\textbf{Note.}} P1 to P15 joined the same younger adults workshop. P16 to P23 joined the same older adult workshop.

\label{tab:participant_demographics}
\end{table*}

The first workshop involved 15 younger adults (P1 to P15), aged 21 to 32 (M = 26.1, SD = 3.3), recruited through a local university. Participants came from varied academic and professional backgrounds, including Human–Computer Interaction, Computer Science, Cyber-security, and Business Management. They represented diverse educational backgrounds, from high school to Ph.D. Younger adults, who have not yet experienced later life stages, could only imagine future needs through aspiration and projection. Their perspectives mainly combined forward-looking imagination with memories of their own childhood, and occasionally with experiences of being around younger relatives or children.

To complement this group, we conducted a second workshop with 8 older adults (P16 to P23), aged 66 to 81 (M = 72.9, SD = 5.2), recruited through senior centers and community networks. While they also had no direct experience with modular robot design, their ideas were grounded in everyday realities of aging, caregiving, and long-term domestic support. In contrast to the younger group’s aspirational orientation, older adults provided reflections informed by firsthand experiences of later life challenges and support needs, and drew on a mix of autobiographical memories and occasionally their experiences as parents or grandparents when reasoning about childhood.

By comparing these two groups under a unified design process, our study combines forward-looking imagination with experience-based reflection. This dual perspective enables us to capture both the speculative possibilities and the grounded constraints of lifespan-oriented modular robots. Demographic details and prior robot experiences of all 23 participants are summarized in Table \ref{tab:participant_demographics}. Beyond these self-reported categories, participants who have experience with robots described prior experiences primarily with
domestic service robots (e.g., robotic vacuum cleaners), educational robots used in classes or labs,
and voice-based smart assistants. Only a small subset mentioned prior exposure to explicitly
“companion-like” robots (e.g., pet-like or socially assistive systems), and those encounters were
typically short and institutional (e.g., museums, care facilities) rather than long-term in the home.

%% Head
\begin{figure*}
\centering
\begin{minipage}[b]{0.55\textwidth}
    \centering
    \includegraphics[width=\linewidth]{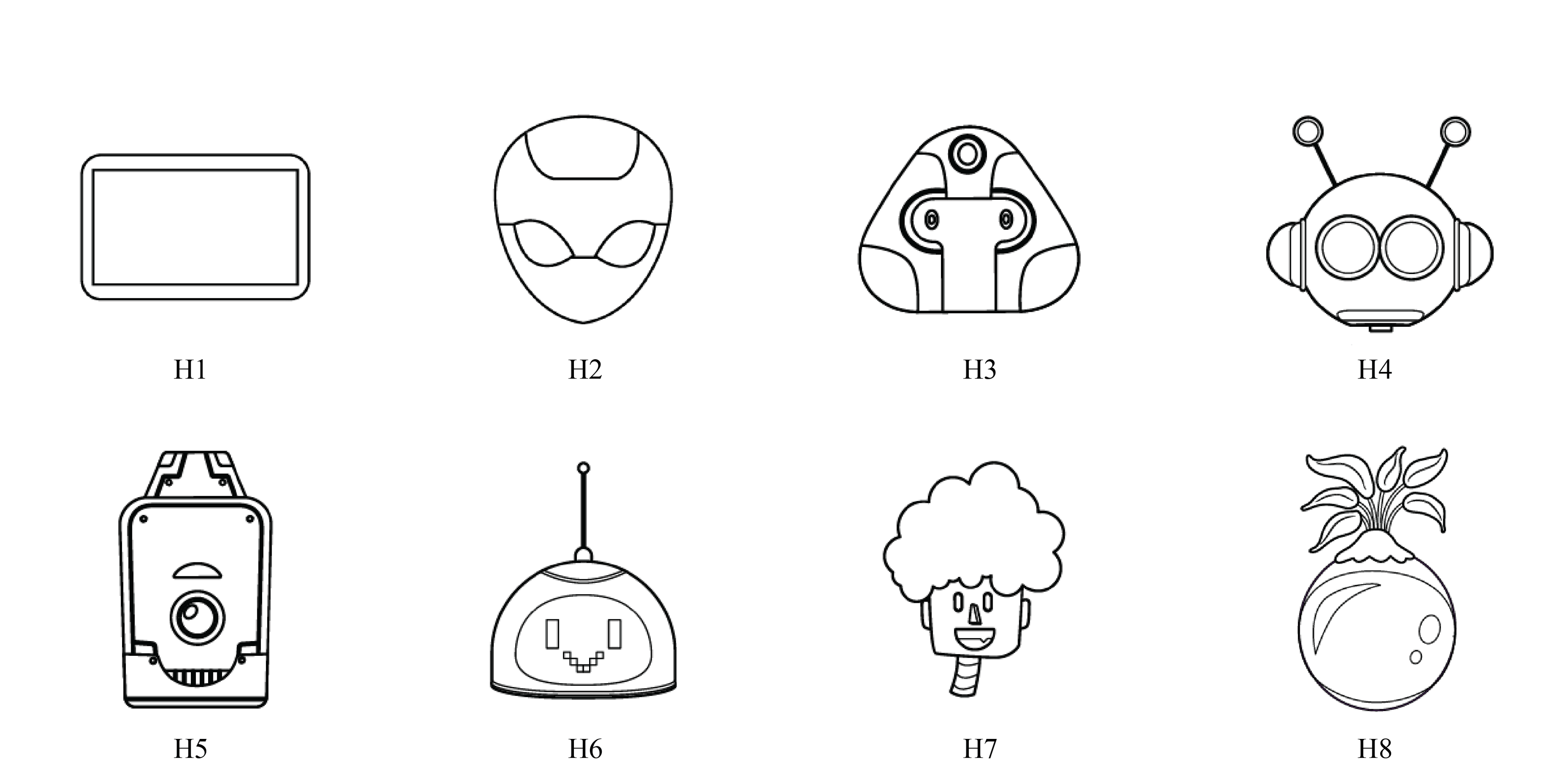}
    \caption{Modular Head (H)}
    \Description{Eight modular robot heads (H1–H8) are depicted in black and white, ranging from a flat tablet screen and simple humanoid faces to dome-shaped, alien-like, and plant-topped heads, illustrating varied visual styles and implied functions.}
    \vspace{3ex} 
    \label{fig:modTH}
\end{minipage}
\hfill
\begin{minipage}[b]{0.42\textwidth}
    \footnotesize
    \centering
    \renewcommand{\arraystretch}{1.3}
    \begin{tabular}{%
    >{\raggedright\arraybackslash}p{0.05\linewidth}%
    >{\raggedright\arraybackslash}p{0.92\linewidth}%
    }
        \specialrule{0.1em}{0em}{0em}
        \multicolumn{2}{c}{\textbf{Potential Function}} \\
        \specialrule{0.1em}{0em}{0em}
        H1 & Tablet screen for data display or expressions. \\
        \specialrule{0.03em}{0em}{0em}
        H2 & Animated expressions, or messages for interaction. \\
        \specialrule{0.03em}{0em}{0em}
        H3 & The triangular shape helps to optimize the movement on unstable terrain. \\
        \specialrule{0.03em}{0em}{0em}
        H4 & Dual antennas can sense and analyze voice, camera-enabled eyes. \\
        \specialrule{0.03em}{0em}{0em}
        H5 & Equipped with 3D projection capabilities. \\
        \specialrule{0.03em}{0em}{0em}
        H6 & Adjusts its expression based on the weather, such as displaying a “sunshine smile” on sunny days. \\
        \specialrule{0.03em}{0em}{0em}
        H7 & Scans the person’s head shape and uses virtual reality to display different hairstyle options. \\
        \specialrule{0.03em}{0em}{0em}
        H8 & Grows and sustains real plants, provides optimal care. \\
        \specialrule{0.1em}{0em}{0em}
    \end{tabular}
    \vspace{3ex} 
    \captionof{table}{Function of Head (H) modules}
    
    \label{tab:funcTH}
\end{minipage}
\end{figure*}
%% Arm
\begin{figure*}
\centering
\begin{minipage}[b]{0.55\textwidth}
    \centering
    \renewcommand{\arraystretch}{1.3}
    \includegraphics[width=\linewidth]{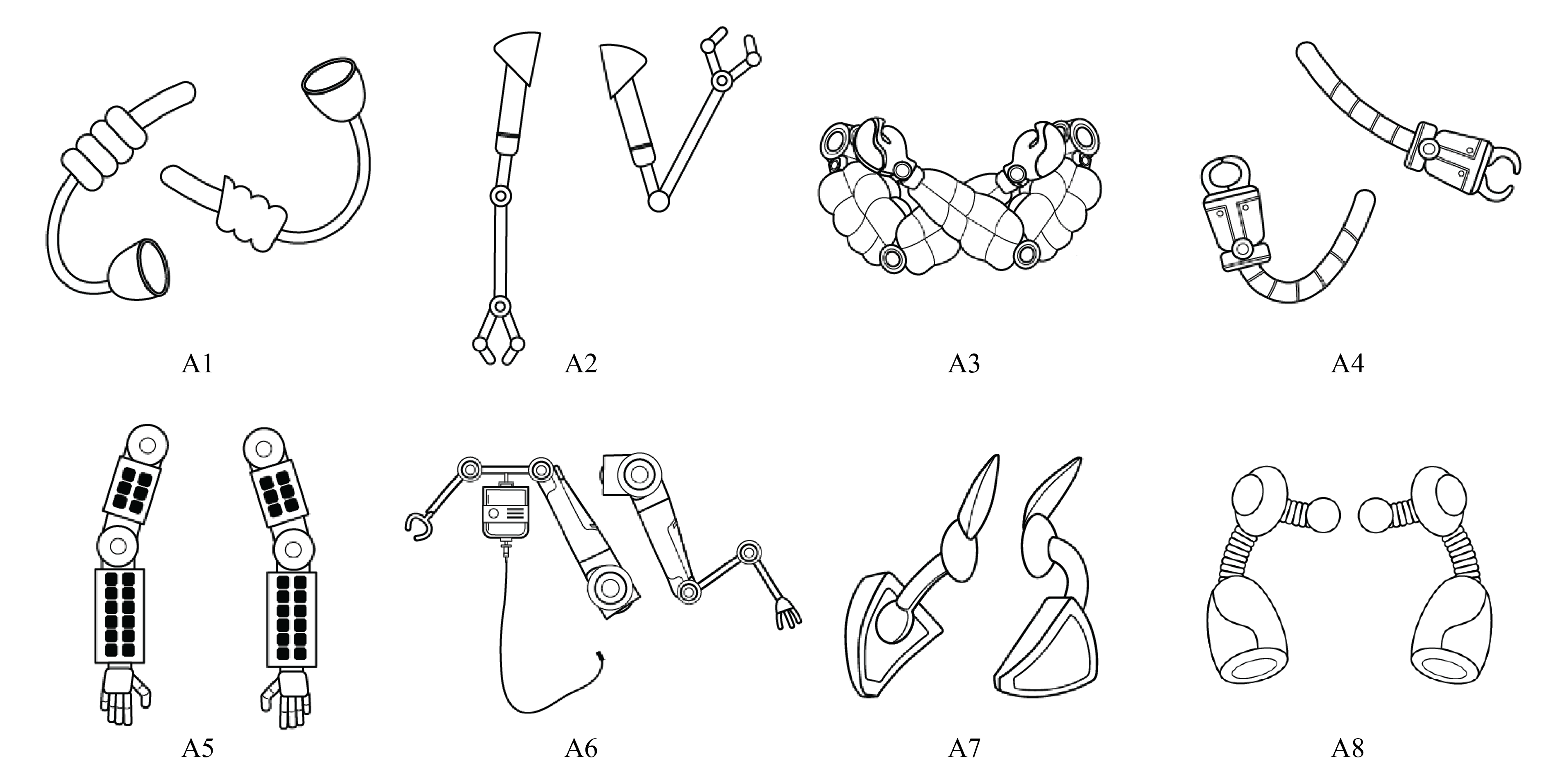}
    \caption{Modular Arm (A)}
    \Description{Eight modular robot arms (A1–A8) are depicted in black and white, including soft cable-like arms, long reach grippers, multi-segment clusters, industrial tool arms, and padded or claw-like limbs to suggest different manipulation and support capabilities.}
    \vspace{3ex} 
    \label{fig:modTA}
\end{minipage}
\hfill
\begin{minipage}[b]{0.42\textwidth}
    \footnotesize
    \centering
    \renewcommand{\arraystretch}{1.3}
    \begin{tabular}{%
    >{\raggedright\arraybackslash}p{0.05\linewidth}%
    >{\raggedright\arraybackslash}p{0.92\linewidth}%
    }
        \specialrule{0.1em}{0em}{0em}
        
        \multicolumn{2}{c}{\textbf{Potential Function}} \\
        \specialrule{0.1em}{0em}{0em}
        
        A1 & Moves like octopus arms, wrapping around objects for enhanced dexterity and control. \\
        \specialrule{0.03em}{0em}{0em}
        A2 & Extendable gripper arms for precision tasks. \\
        \specialrule{0.03em}{0em}{0em}
        A3 & Gel-cushioned arms perform comforting gestures like hugs. \\
        \specialrule{0.03em}{0em}{0em}
        A4 & Suitable for flexible movement in confined spaces. \\
        \specialrule{0.03em}{0em}{0em}
        A5 & Solar panel arms for energy harvesting. \\
        \specialrule{0.03em}{0em}{0em}
        A6 & Suitable for hospitals or home care, which can hold items such as IV bags or deliver medication. \\
        \specialrule{0.03em}{0em}{0em}
        A7 & Integrated with protective materials to provide shielding or absorb impacts. \\
        \specialrule{0.03em}{0em}{0em}
        A8 & Controls metal objects without physical contact with built-in spring mechanisms that allow them to extend and adapt to various tasks. \\
        \specialrule{0.1em}{0em}{0em}
    \end{tabular}
    \vspace{3ex} 
    \captionof{table}{Function of Arm (A) modules}
    \label{tab:funcTA}
\end{minipage}
\end{figure*}
%% Upper Body
\begin{figure*}
\centering
\begin{minipage}[b]{0.55\textwidth}
    \centering
    \includegraphics[width=\linewidth]{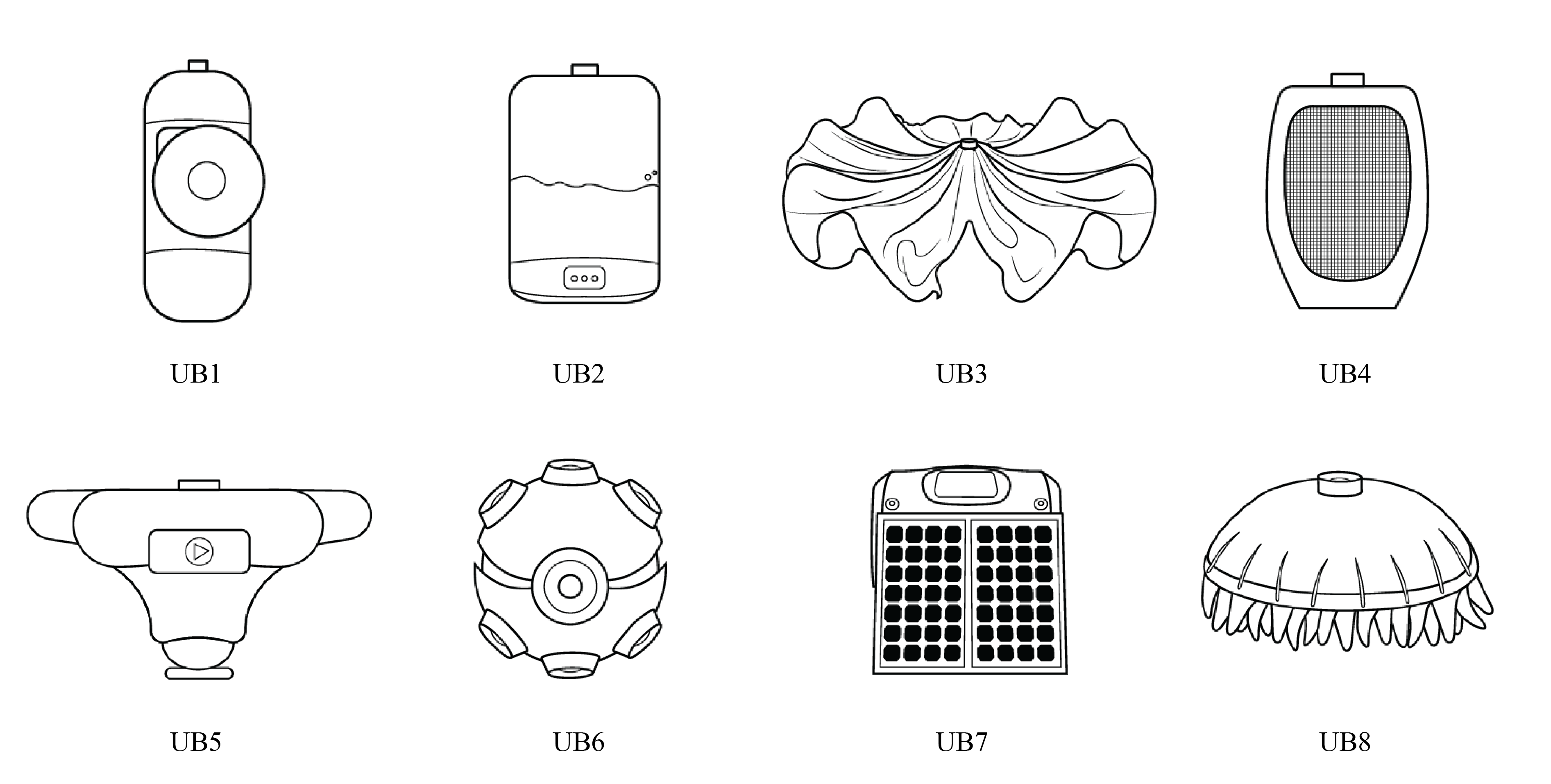}
    \caption{Modular Upper Body (UB)}
    \Description{Eight modular upper bodies (UB1–UB8) are depicted in black and white, including cylindrical tanks, draped fabric-like torsos, flat panels, hubs with multiple attachment points, and domed cushion shapes, offering different options for carrying, protection, and interaction surfaces.}
    \vspace{3ex} 
    \label{fig:modTUB}
\end{minipage}
\hfill
\begin{minipage}[b]{0.42\textwidth}
    \footnotesize
    \centering
    \renewcommand{\arraystretch}{1.3}
    \begin{tabular}{%
    >{\raggedright\arraybackslash}p{0.05\linewidth}%
    >{\raggedright\arraybackslash}p{0.92\linewidth}%
    }
        \specialrule{0.1em}{0em}{0em}
        
        \multicolumn{2}{c}{\textbf{Potential Function}} \\
        \specialrule{0.1em}{0em}{0em}
        
        UB1 & Equipped with an advanced rotating camera that provides observation for security and exploration. \\
        \specialrule{0.03em}{0em}{0em}
        UB2 & Aromatherapy diffuser releases calming scents and music. \\
        \specialrule{0.03em}{0em}{0em}
        UB3 & Fabric body for adaptive protection. \\
        \specialrule{0.03em}{0em}{0em}
        UB4 & Generates soundscapes and voice responses to enhance human-robot communication. \\
        \specialrule{0.03em}{0em}{0em}
        UB5 & Plays media and displays messages. \\
        \specialrule{0.03em}{0em}{0em}
        UB6 & Gathers data from harsh and extreme environments for exploration. \\
        \specialrule{0.03em}{0em}{0em}
        UB7 & Equipped with solar panels for energy harvesting. \\
        \specialrule{0.03em}{0em}{0em}
        UB8 & Provides a soft and gentle touch to enhance comfort. \\
        \specialrule{0.1em}{0em}{0em}
    \end{tabular}
    \vspace{3ex} 
    \captionof{table}{Function of Upper Body (UB) modules}
    \label{tab:funcTUB}
\end{minipage}
\end{figure*}
\begin{figure*}
\centering
\begin{minipage}[b]{0.55\textwidth}
    \centering
    \includegraphics[width=\linewidth]{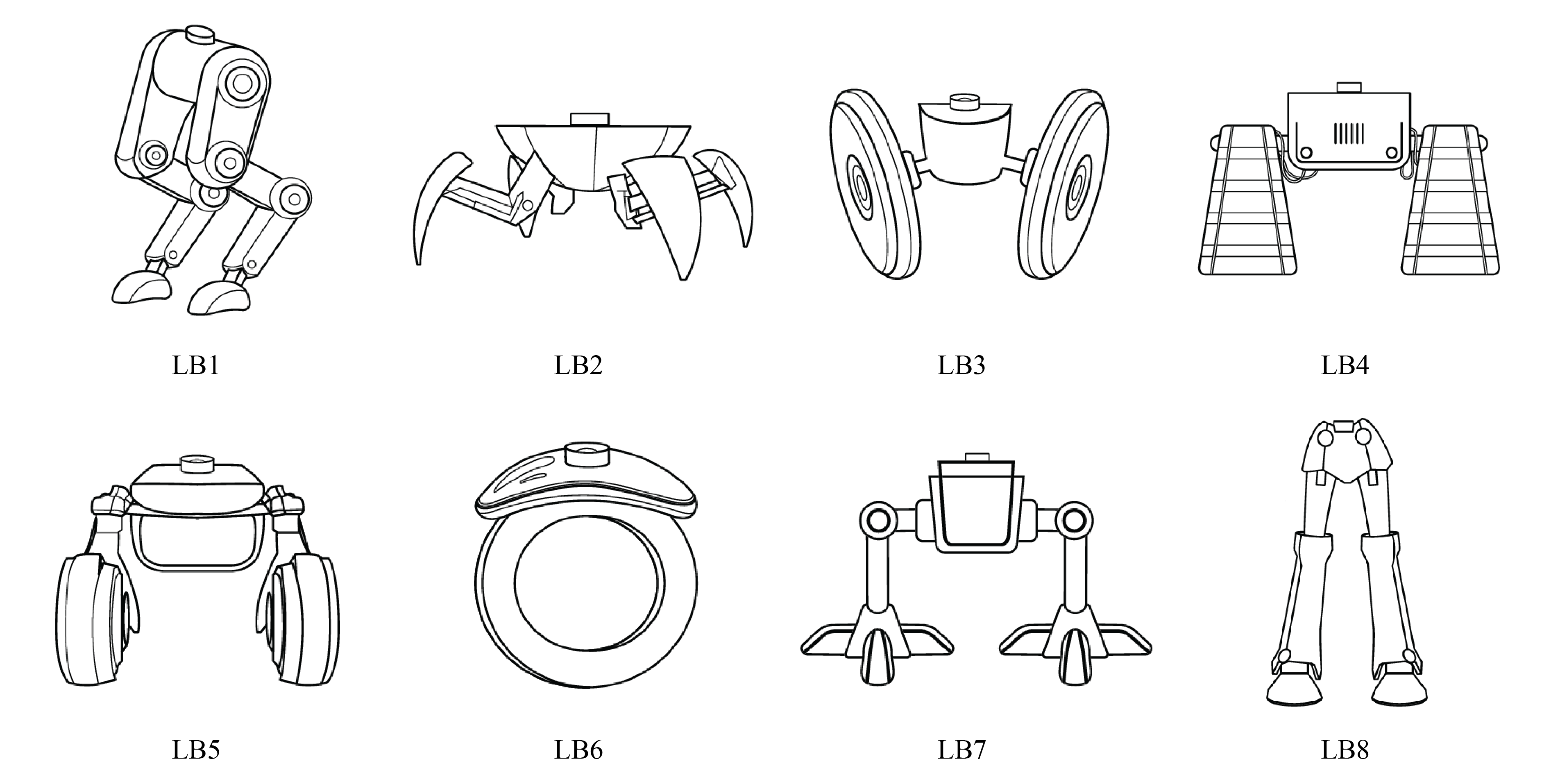}
    \caption{Modular Lower Body (LB)}
    \Description{Eight modular lower-body (LB1—LB8) are depicted in black and white, including different locomotion options such as bird-like legs, spider legs, wheeled bases, and wheelchair-like platforms.}
    \vspace{3ex} 
    \label{fig:modTLB}
\end{minipage}
\hfill
\begin{minipage}[b]{0.42\textwidth}
    \footnotesize
    \centering
    \renewcommand{\arraystretch}{1.3}
    \begin{tabular}{%
    >{\raggedright\arraybackslash}p{0.05\linewidth}%
    >{\raggedright\arraybackslash}p{0.92\linewidth}%
    }
        \specialrule{0.1em}{0em}{0em}
        
        \multicolumn{2}{c}{\textbf{Potential Function}} \\
        \specialrule{0.1em}{0em}{0em}
        
        LB1 & Bird legs for jumping mobility. \\
        \specialrule{0.03em}{0em}{0em}
        LB2 & Spider legs for climbing and stability. \\
        \specialrule{0.03em}{0em}{0em}
        LB3 & Wheeled base for fast movement. \\
        \specialrule{0.03em}{0em}{0em}
        LB4 & Transforms between walking, rolling, or hovering based on the environment’s demands. \\
        \specialrule{0.03em}{0em}{0em}
        LB5 & Wheelchair-like design for adaptive mobility across various terrains, providing comfort and effortless movement. \\
        \specialrule{0.03em}{0em}{0em}
        LB6 & A fully rotational wheel that allows to move smoothly. \\
        \specialrule{0.03em}{0em}{0em}
        LB7 & Digs into soft terrain like sand or soil to secure its position. \\
        \specialrule{0.03em}{0em}{0em}
        LB8 & Humanoid legs for walking and balance. \\
        \specialrule{0.1em}{0em}{0em}
    \end{tabular}
    \vspace{3ex} 
    \captionof{table}{Function of Lower Body (LB) modules}
    \label{tab:funcTLB}
\end{minipage}
\end{figure*}

\subsection{Workshop Set-up} \label{set}

We conducted a 90-minute co-design workshop structured into three stages: introduction, hands-on design, and reflection.

\begin{figure*}[t]
    \centering
    \includegraphics[alt={a group of younger participants seated around a table engaged in the study activities}, width=0.5\linewidth]{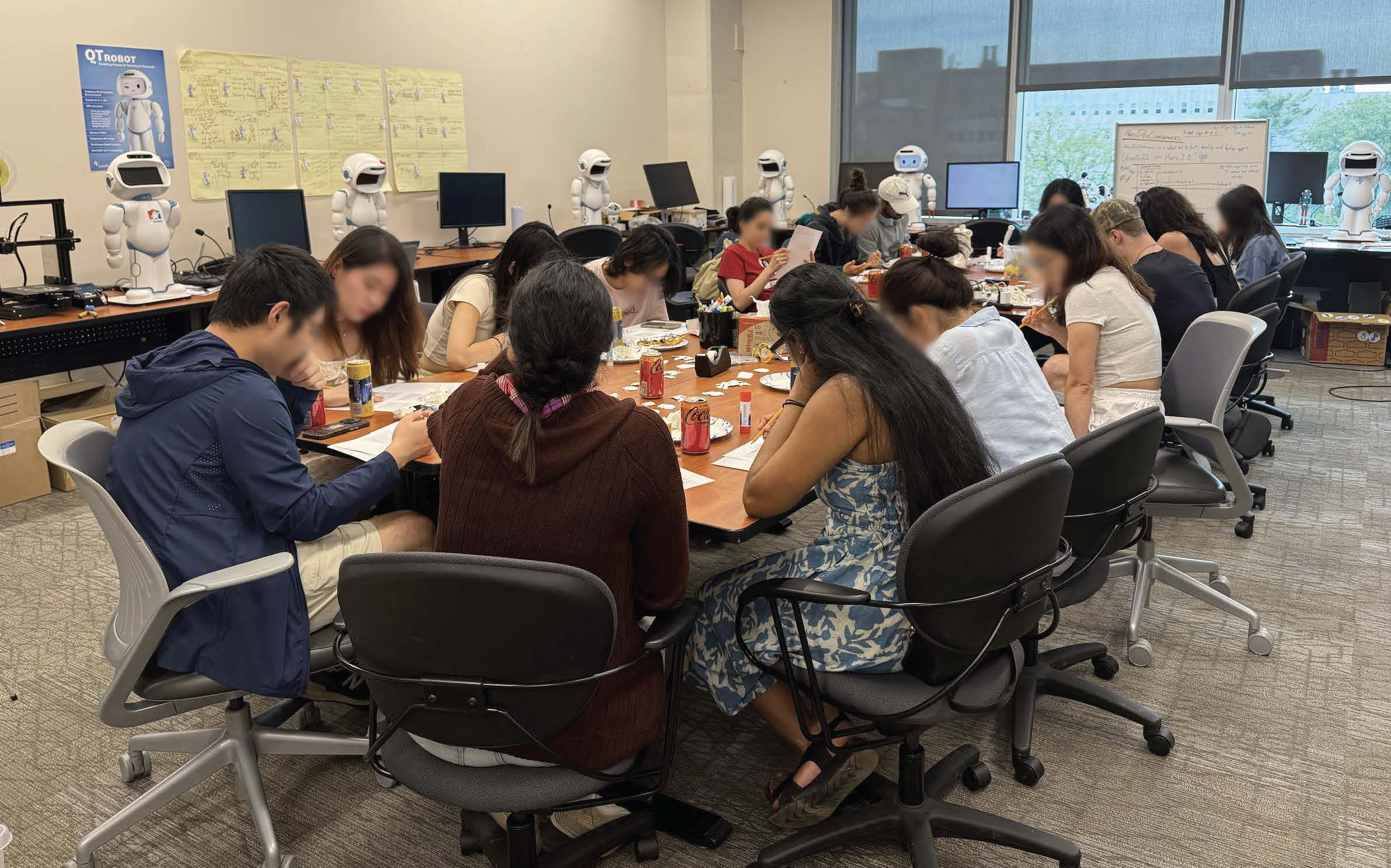}
    \caption{Workshop 1 with 15 younger adults}
    \Description{Photograph of Workshop 1 with fifteen younger adults, showing participants seated around a large table in a classroom, sketching and assembling modular robot designs with provided materials.}
    \label{fig:ws1}
\end{figure*}

\begin{figure*}[t]
    \centering
    \includegraphics[width=0.5\linewidth]{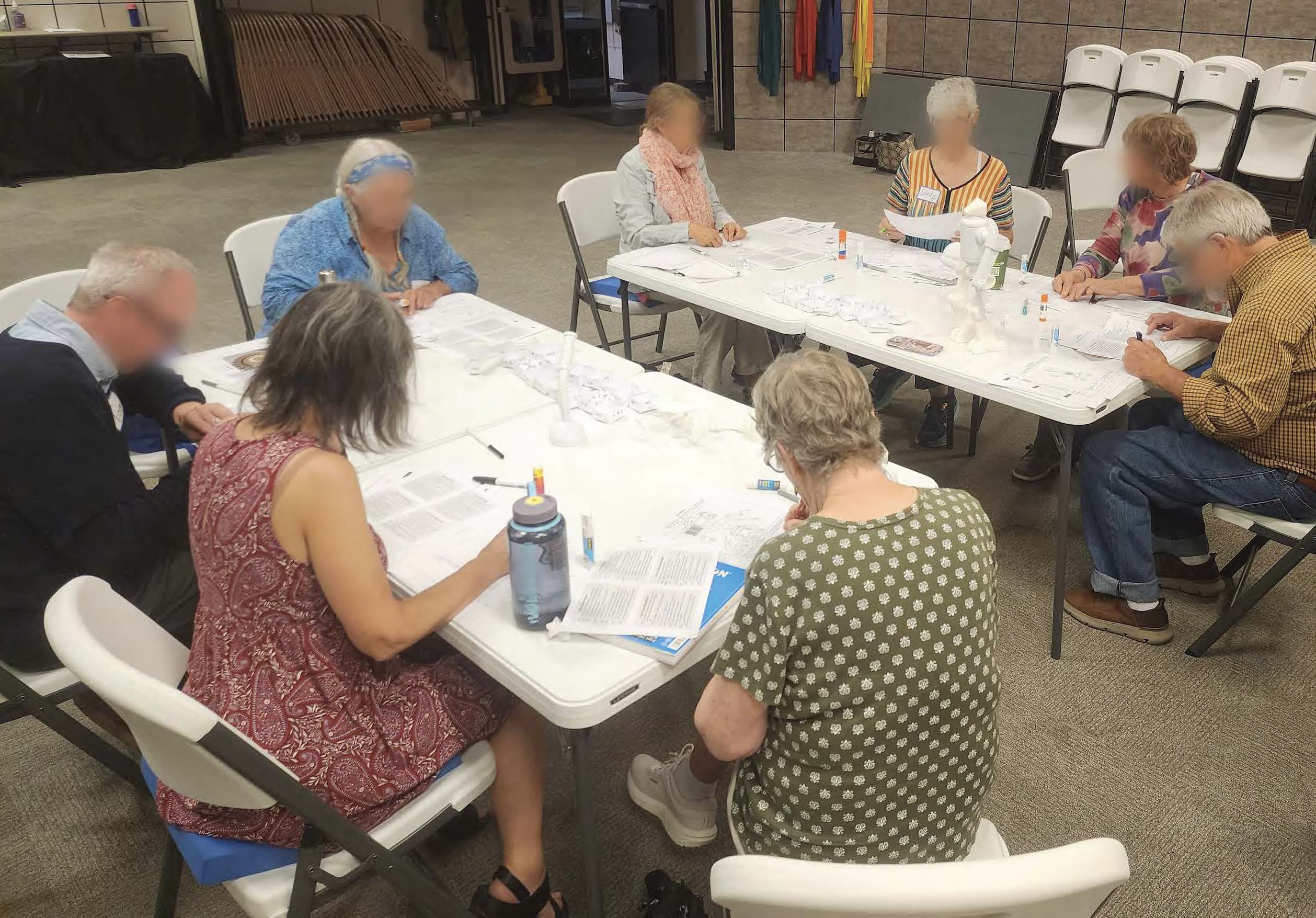}
    \caption{Workshop 2 with 8 older adults}
    \Description{Photograph of Workshop 2 with eight older adults, showing participants seated around a rectangular table in a community room, engaged in drawing and discussing modular robot designs.}
    \label{fig:ws2}
\end{figure*}

\textbf{Introduction.} At the beginning, we introduced the idea of modular robots to the participants. We used the module design set and its associated tables of potential functions from \citeauthor{chen2025sustainable} \cite{chen2025sustainable} as the shared design basis for the participants to assemble or sketch their designed robot. 
Heads (H), arms (A), an upper body (UB), and a lower body (LB) served as a basic structural scaffold, but participants were not required to use all of them. To keep the focus on personal needs and functions rather than appearance, participants first selected desired functions from text-only tables (e.g., expression generation, climbing mobility, energy harvesting) without showing images (see Table \ref{tab:funcTH}, Table \ref{tab:funcTA}, Table \ref{tab:funcTUB}, Table \ref{tab:funcTLB}). Only after they had made these choices did we reveal the corresponding visual module designs and invite them to assemble or sketch their robots. This sequence encouraged participants to imagine one robot that could be reconfigured to support different life stages, rather than designing separate robots for each age. \label{instrument}

\textbf{Hands-on design.} Next, participants considered three major life stages, childhood, adulthood, and older adulthood, and were asked to design a modular robot for themselves. They imagined how the same robot might assist and accompany themselves rather than designing for an abstract or generic user. In doing so, participants chose whether to speak from the vantage point of their younger self, their current adult self, or their anticipated older self. They speculated on how a modular robot might support their everyday lives across these stages. We deliberately framed the task around home and nearby everyday settings rather than industrial or large-scale public environments. Within this framing, participants were free to assign any roles they found meaningful, ranging from tool-like helpers and task-oriented assistants to more social or emotionally supportive partners. To emphasize continuity, they were asked to imagine how the \textit{same robot} could gradually change its form by replacing or adding modules across life stages. For each life stage, they selected relevant functions, retrieved corresponding stickers of modules, and assembled their envisioned robots using stickers on a template (see \autoref{fig:modTH}, \autoref{fig:modTA}, \autoref{fig:modTUB}, \autoref{fig:modTLB}). If the provided modules did not fit their intentions, they were encouraged to sketch their own, fostering creativity and revealing unmet design needs. 

\textbf{Reflection and sharing.} Finally, participants presented their designs in small groups. They described how the same robot might transform over time to support changing needs, which modules would remain constant, and which would be added, removed, or replaced. Discussions highlighted how modularity made it possible for one robot to serve as a lifelong companion, rather than a single-purpose or age-specific device. Participants reflected on which parts felt essential, unnecessary, or emotionally significant, offering insights into how embodiment and continuity shape human–robot relationships.

\textbf{Outcomes.} By centering user imagination and emphasizing continuity through modularity, the workshop revealed a multiplicity of robotic forms. Designs ranged from fully-bodied companions to minimalist robots consisting only of a head or torso. These results challenge the assumption that robots must always be anthropomorphic, instead of surfacing pluralistic visions of domestic technology that evolve across the human lifespan.

\subsection{Data Collection and Analysis}

All workshop sessions were audio-recorded with participant consent in their entirety for later reference and transcription. The primary data also includes the modular robot designs that participants created using provided stickers, supplemented by hand-drawn sketches and written annotations made during the workshop. Each participant produced three distinct robot designs, corresponding to childhood, adulthood, and older adulthood. These physical artifacts were collected and digitized as high-resolution PDF scans immediately following each session, ensuring that the design details and annotations were preserved for subsequent analysis.

In addition to the artifacts, we took detailed field notes during the workshops to capture participants’ verbal explanations, spontaneous comments, and group-level discussions. Participants were encouraged to annotate their designs with brief descriptions of intended functions and rationales for their module selections. These annotations, together with their verbal reflections, provided valuable contextual information that helped us interpret not only what functions participants prioritized but also why these functions mattered to them at different life stages. The combination of multimodal data—visual artifacts, textual annotations, spoken accounts, and researcher notes—allowed us to build a rich dataset that encompassed both the tangible outcomes of design activities and the reasoning processes behind them.

We adopted an interpretive qualitative approach \cite{adams2008qualititative, boehner2007hci}, integrating transcripts, artifacts, and annotations into a single analytic corpus. Two researchers first reviewed a subset of data to generate initial coding categories around robot functions, embodiment choices, and life-stage expectations. We then engaged in iterative open coding and constant comparison, clustering codes into broader categories that captured both convergent patterns and divergent perspectives. Building on this, we applied thematic analysis \cite{clarke2014thematic, levinson2024bow}, a widely used method within HCI’s qualitative research tradition \cite{fan2025user,cooper2022systematic,xiao2025let}, to identify underlying patterns across the dataset. Through multiple rounds of team discussions, we refined these categories into three overarching themes, which capture not only functional expectations and design preferences but also participants’ values and concerns about long-term human–robot relationships. To ensure credibility, we revisited raw data to ground emerging themes and documented our interpretive decisions through analytic memos.

\subsection{Ethics}

All study procedures were reviewed and approved by our university’s Institutional Review Board (IRB). Before each workshop began, participants were provided with an informed consent form that described the purpose of study, procedures, potential risks, and their right to withdraw at any time without penalty. Written consent was obtained from all participants prior to participation. 

To protect participants' privacy, all identifying information was removed from the dataset. Audio recordings were transcribed and anonymized, and design artifacts were labeled only with participant codes (e.g., P1–P23). Data were stored securely and used solely for research purposes. In reporting our findings, we use pseudonyms or participant identifiers to ensure confidentiality.

\section{Findings}
%\ref{fig:drawing01}, \ref{fig:drawing02} and \ref{fig:drawing03} 
We present a total of 65 sketches produced across the two workshops: 45 sketches from 15 participants in Workshop 1, and 20 sketches from 8 participants in Workshop 2. The complete collection of sketches is shown in Appendix ~\ref{appendix:all_sketches}. In the main text, we focus on selected, annotated examples to illustrate how participants used modular configurations to realize personalization, adaptability, and sustainability. It was noted that P22$_{\text{66/F}}$ preferred a stable look and therefore kept the same design across childhood, adulthood, and older adulthood. P23$_{\text{81/F}}$ only completed the older adulthood stage due to time constraints.

Our study highlights that modular design functions not only as a technical mechanism but as a design strategy for enabling personalization, adaptability, and sustainability. These qualities often appeared interlinked in participants’ designs and reflections, rather than existing as isolated concerns. In the following sections, we present three dimensions where modularity played a critical role in shaping participants’ visions of robots across life stages. 

\begin{figure*}[t]
    \centering
    \includegraphics[width=1\linewidth]{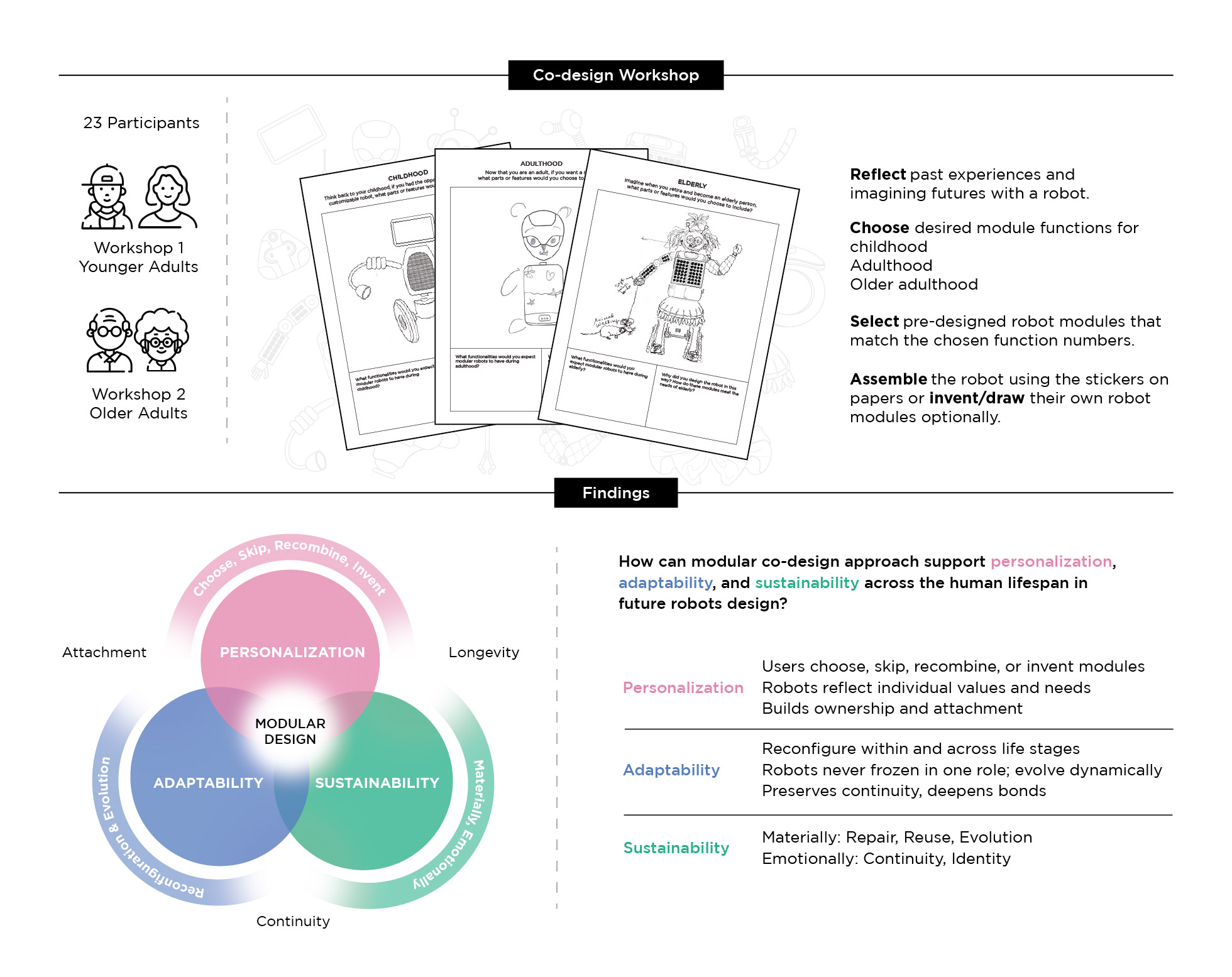}
    \caption{Overview of the Study: Participants from two age groups reflected on life stages, selected robot functions, and assembled modular designs. Findings show how modular design supports personalization, adaptability, and sustainability across the lifespan.}
    \Description{Overview of the study design, depicting the two co-design workshops (younger and older adults), the three workshop phases (introduction, hands-on design, reflection), and how these activities feed into the analysis of personalization, adaptability, and sustainability.}
    \label{fig:placeholder}
\end{figure*}

\subsection{Personalization through Modular Design}

Modular design allowed participants to treat robots as creative canvases, enabling deep personalization. Participants framed this as essential for maintaining long-term relevance and personal connection across life stage. Rather than simply “liking customization,” they used modularity to negotiate whether a robot still felt appropriately matched to their changing roles and self-image at each life stage. Participants could selectively choose, skip, remix, or even invent their own modules to construct robots that reflected their individual priorities. These priorities reflect a desire for robots that can grow and transform alongside their users, without losing their core identity.  

Across our findings, in particular, three dimensions of personalization became apparent. First, participants expressed \textit{life-stage demands} by reshaping a single robot into different roles, such as playmate in childhood, assistant in adulthood, and caregiver in older adulthood, demonstrating how modularity could align shared needs with highly individualized visions (see \autoref{per1}). Second, they personalized robots through \textit{choice and reconfiguration}, not only adding or omitting modules but also rearranging them in unconventional ways, showing that agency lies as much in what is excluded or repurposed as in what is added (see \autoref{per2}). Third, they pursued \textit{customization beyond provided parts}, sketching new modules, hybrid forms, and symbolic features, extending robots into carriers of emotional meaning and personal storytelling (see \autoref{per3}). These strategies illustrate how modular design supports personalization as an active, creative process—enabling robots to remain meaningful across different contexts, stages of life, and individual imaginations.

\subsubsection{Personalization for Distinct Life-Stage Demands} \label{per1}

\begin{figure*}[t]
    \centering
    \includegraphics[width=0.5\linewidth]{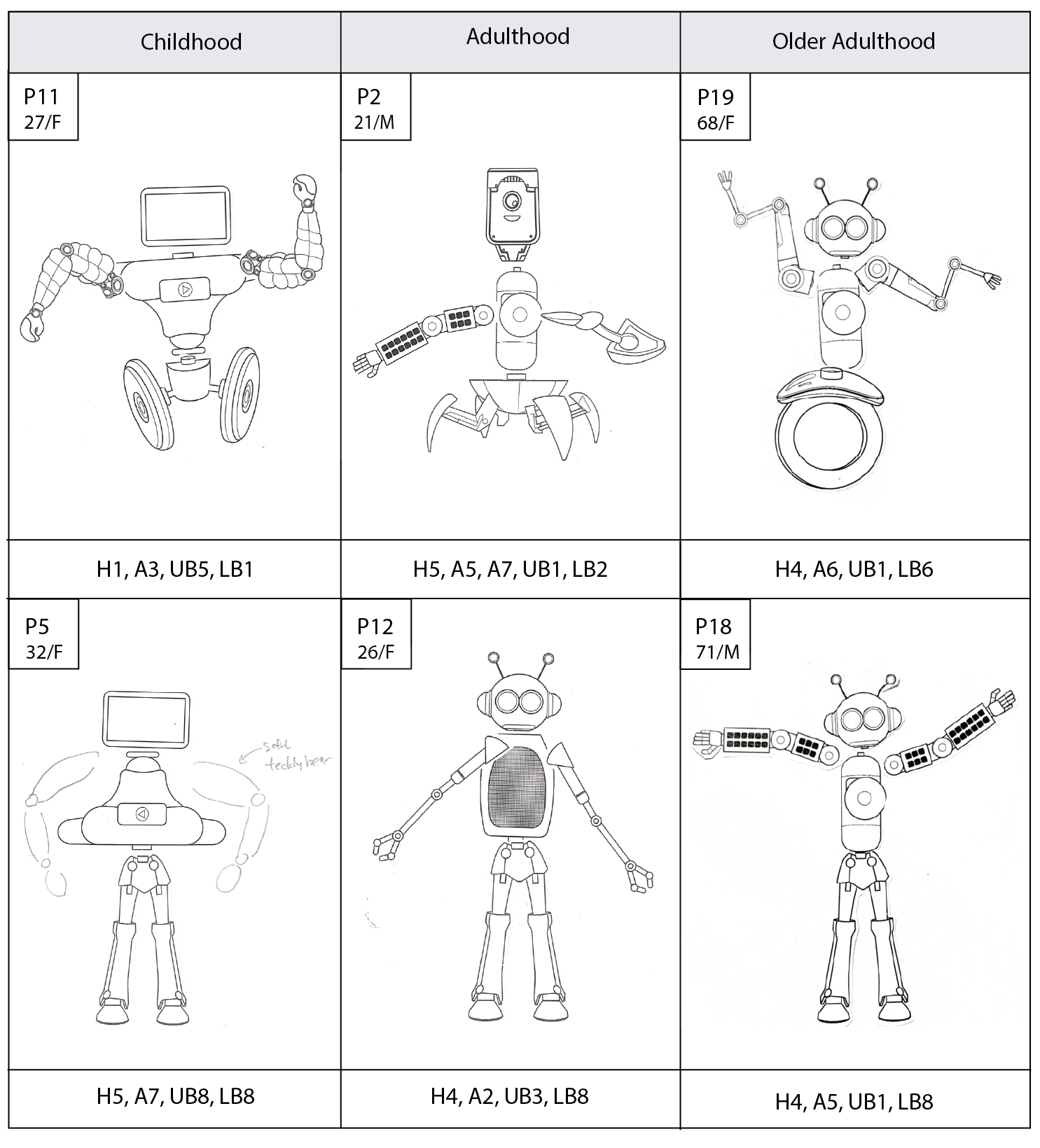}
    \caption{Examples of PAS design combinations by the participants for different life stages}\label{fig:personalization}
    \Description{A 2×3 grid of participant sketches showing different modular robot configurations for childhood, adulthood, and older adulthood, illustrating how the same parts are recombined to meet changing life-stage needs.}
\end{figure*}
All participants in two co-design workshops consistently imagined modular robots as companions that could evolve with them, but the roles and functions varied across life stages. Childhood designs centered on play, exploration, and protection; adulthood designs shifted toward productivity, household balance, and affective support; and older adulthood designs emphasized mobility, stability, and companionship. Across these transitions, modularity was valued as the mechanism that enabled personalization, allowing a single robot to be continually reshaped to reflect changing priorities and intimate preferences over the lifespan.

\textbf{Childhood.} In childhood, participants consistently emphasized play, imagination, and protection, which are characteristics of this formative stage. Robots were imagined as companions that could both entertain and safeguard, reflecting the dual role of childhood as a time of exploration and vulnerability. For some, play meant media and games: P4$_{\text{27/F}}$ envisioned a tablet head to play cartoons, gel-cushioned arms for hugging, and legs for park play, explaining: \textit{“As a child, a doll that can jump, play music, images to play around. Tablet head to play cartoon, soft arm to hug, body for protection, leg to play in parks.”} Similarly, P5$_{\text{32/F}}$ and P11$_{\text{27/F}}$ highlighted entertainment functions like YouTube and chasing games, positioning robots as sources of fun and distraction. Others expanded play into social, peer-like interaction: P17$_{\text{73/M}}$ asked for a robot to \textit{“play card games with me and even jump up if it wins, have human expressions to understand me and see me,”} selecting expressive heads and gripper arms to create an engaging partner. Education and safety were equally central, showing how childhood designs merged curiosity with protection. P2$_{\text{21/M}}$ imagined a robot that should \textit{“be able to teach and educate the child,”} equipping it with instructional modules, while P10$_{\text{29/F}}$ emphasized outdoor safety, envisioning a \textit{“camera-based robot [that] can detect dangerous environment.”} For P1$_{\text{24/M}}$, protection was paramount: \textit{“Something to protect me from getting hurt.”} These examples illustrate how, despite shared needs for play, learning, and safety, each participant translated these into unique module combinations, such as educator, playmate, or protector, demonstrating how modularity enabled personalized visions of childhood companionship.

\textbf{Adulthood.} In adulthood, participants’ expectations shifted toward balancing productivity, domestic management, and emotional relief, reflecting the pressures and responsibilities of this stage of life. Robots were framed as both efficient assistants and affective companions. For workplace productivity, modularity enabled targeted tools: P2$_{\text{21/M}}$ equipped his design with a projection head to support conferences, while P9$_{\text{24/F}}$ stressed efficiency: \textit{“I think I value efficiency and intelligence help the most in my adulthood. I am hoping for something that helps me in my career and fasten my daily tasks.”} Others focused on household support, recognizing the strain of combining work and home responsibilities. P7$_{\text{28/F}}$ described a robot that could \textit{“help with any kind of chores, like cooking, cleaning up, just be supportive when I need it in my busy life.”} Similarly, P10$_{\text{29/F}}$ envisioned a multitasking helper to take over laundry and vacuuming. At the same time, participants underscored emotional needs. P14$_{\text{26/F}}$ articulated a desire for robots to replicate familial intimacy: \textit{“If you could set the hugs like the way your mom hugs you, it can mimic the same hugs.”} Here, modular arms were not merely functional but embodied personalized forms of affection. These accounts show how adults commonly valued efficiency and support, yet each participant emphasized different priorities: career acceleration, household relief, or emotional reassurance. Through modularity, a single adult-stage robot could be reconfigured into distinct forms, embodying the principle of personalization by aligning with diverse but intersecting adult demands.

\textbf{Older adulthood.} In later life, participants highlighted needs tied to aging, like mobility, stability, caregiving, and companionship, while also seeking dignity and continuity. Mobility was repeatedly mentioned, but interpreted in different ways. P17$_{\text{73/M}}$ emphasized autonomy, selecting a wheelchair-like base to preserve independence, while P19$_{\text{68/F}}$ stressed practical support, imagining a robot to \textit{“retrieve things within the home.”} Emotional presence was equally important. P18$_{\text{71/M}}$ designed a walking partner that could \textit{“move at my pace, slower and faster, explore the environment and share its observations,”} turning a mobility aid into a shared companion for daily life. Others underscored comfort and caregiving. P9$_{\text{24/F}}$ imagined retirement supported by softness and reassurance, explaining: \textit{“I just need comfort and still efficient help (especially for help with health and connection with family).”} By configuring modules into plush, supportive forms, she designed a robot that blended efficiency with emotional comfort. These designs demonstrate both shared priorities in older adulthood, like mobility, companionship, and safety, and diverse forms of personalization of how to achieve them. For some, independence was central; for others, gentle presence or caregiving mattered more. Modularity allowed these differences to be realized within a shared framework, making the robot simultaneously a symbol of collective needs in older adulthood and a personalized expression of individual values and circumstances.

These cases across different life stages demonstrate how modularity transformed robots into deeply personal companions rather than one-size-fits-all devices. Even when participants shared common needs, such as play, efficiency, or care, each translated them into distinct robotic forms that reflected their own values and life circumstances. Personalization thus emerged as the central contribution of modular design, enabling robots to adapt across life stages while remaining meaningfully aligned with individual imaginations and experiences.

\subsubsection{Choice and Reconfiguration} \label{per2}
\begin{figure*}[t]
    \centering
    \includegraphics[width=0.5\linewidth]{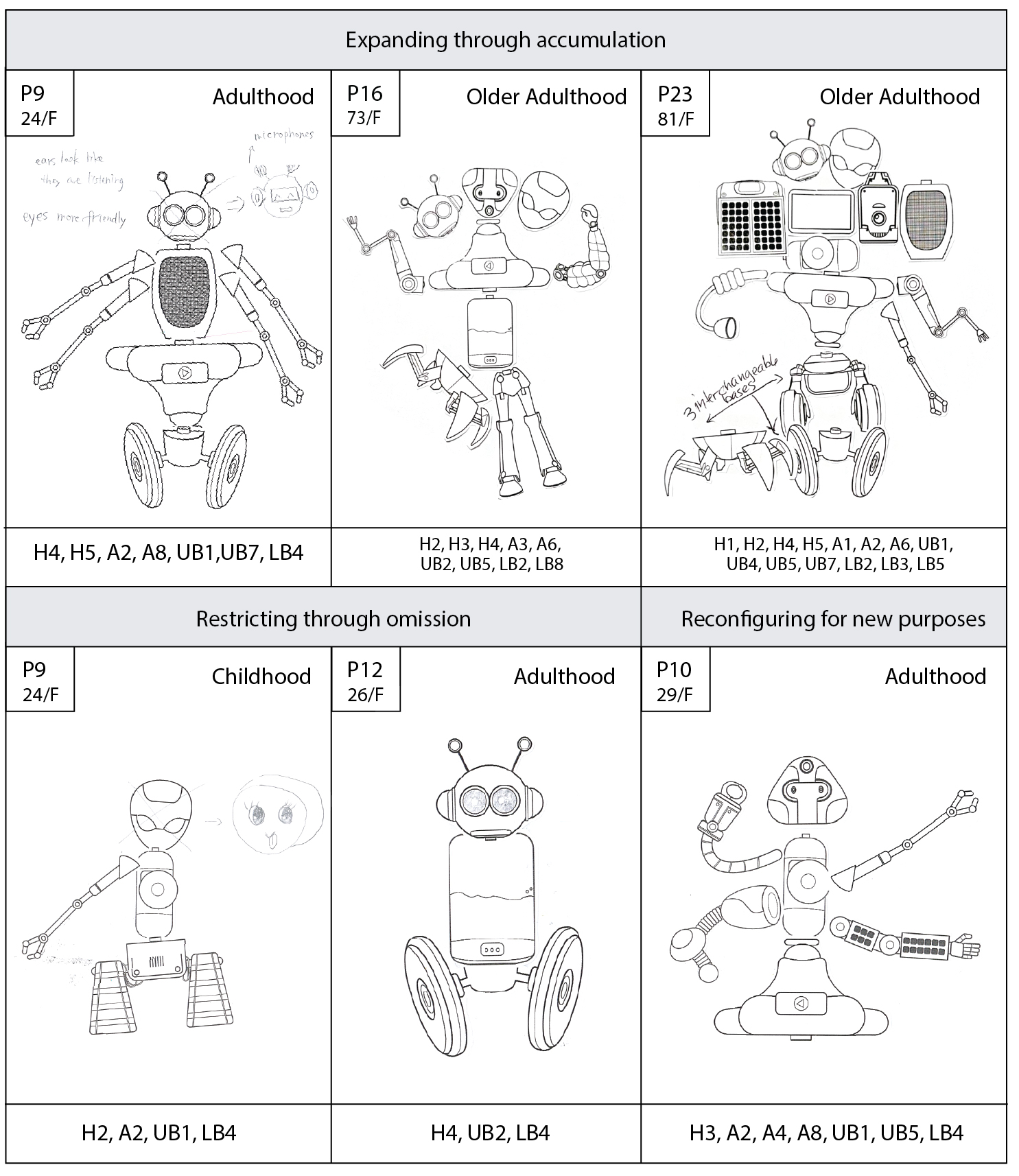}
    \caption{Choice and reconfiguration examples by the study participants}\label{fig:figreconfig}
    \Description{A 2×3 grid of participant sketches showing three robots expanded with extra modules, two robots simplified by omitting parts, and one robot with modules rearranged, highlighting accumulation, omission, and reconfiguration as personalization strategies.}
\end{figure*}
A central way modularity supported personalization in our study was by granting participants freedom in both choice and reconfiguration. Rather than being asked to build a “correct” or “complete” robot, participants were encouraged to treat modules as open design materials. This speculative framing gave them permission to question default assumptions about robot form and to imagine alternatives that reflected their own priorities. Figure \ref{fig:figreconfig} shows examples of designs with redundant parts, certain bodyparts omitted, or reconfigurations of parts for other purposes.

\textbf{Expanding through accumulation.} Many participants exercised personalization by adding more modules than the standard four-part template suggested, building larger or more complex robots to serve multiple functions. Younger adults in our workshops often added extra arms to support multitasking, while older adults expanded the overall configuration to address physical or care needs. For example, P9$_{\text{24/F}}$ in adulthood envisioned her robot as a learning and task-support partner, equipping it with four arms (A2) for precision activities while maintaining mobility (LB3). She explained: \textit{“4 arms help me with work, more intelligent, move fast enough.”} Similarly, P16$_{\text{73/F}}$ layered multiple heads and arms, describing her older adulthood design as a care assistant capable of addressing diverse needs. \textit{“Lots of mobility to help me get around anywhere. I will want to keep experiencing independence and adventure like I did as a child.”} She selected LB2/LB8 for stability, A2/A3/A6 for assistance, and H2/H3/H4 for sensing and communication. P23$_{\text{81/F}}$ also envisioned redundancy and flexibility, noting: \textit{“for elderly people, the robot has to be very stable and help keep them steady during the day.”} She combined LB2 (spider legs), LB3 (wheeled base), and LB5 (wheelchair-like mobility), explicitly emphasizing \textit{“3 interchangeable bases.”} These examples show how participants expanded configurations to reflect complex priorities, from efficiency in adulthood to stability and independence in later life.

\textbf{Restricting through omission.} In contrast, some participants personalized robots by deliberately omitting modules. This approach highlighted different philosophies of support, particularly in childhood designs. For instance, P9$_{\text{24/F}}$ imagined a childhood robot that could help satisfy curiosity without getting “hands dirty,” using A2 arms to fetch objects. Yet P12$_{\text{26/M}}$ resisted such assistance, avoiding arms altogether: \textit{“I chose not to add soft arms because I believe children should have the opportunity to explore the world independently. If the robot has arms, it feels as though the robot is doing things for the child, not with a child.”} He carried this stance into adulthood, again omitting arms. He configured H4 (sensor head), UB2 (storage body), and LB4 (wheeled base) to prioritize awareness and mobility over manipulation, arguing that the robot should support independence rather than perform tasks on his behalf. By leaving out modules, participants could express values of independence or restraint, showing that personalization was not always about “more,” but about aligning designs with personal beliefs about development and care.

\textbf{Reconfiguring for new purposes.} Beyond selection and omission, some participants reconfigured modules in unconventional ways, challenging the spatial logic of the system. Rather than treating modules as fixed anatomical parts, they rearranged them to achieve new functions. For example, P10$_{\text{29/F}}$ placed an upper-body module in the lower-body position, flipping the hierarchy to repurpose it as a base. She explained: \textit{“It will equip several arms for various tasks.”} This adjustment created a more stable robot capable of supporting multiple arms, reimagining the system as a sturdier household assistant for domestic chores. Such creative reconfigurations illustrate how participants saw modules not as rigid parts but as flexible design elements open to reinterpretation.

\textbf{From choice to co-creation.} These strategies, accumulation, omission, and reconfiguration, show how modularity enabled personalization beyond surface-level customization. Participants exercised freedom to decide what to include, what to exclude, and how to rearrange modules, producing robots that aligned with their own needs, values, and life-stage demands. In this framing, modular design elevates personalization into an active process of co-creation, where users define whether the robot should act as a helper, a peer, or a supportive background presence.

\subsubsection{Customization Beyond Provided Parts} \label{per3}
\begin{figure*}[t]
    \centering
    \includegraphics[width=0.5\linewidth]{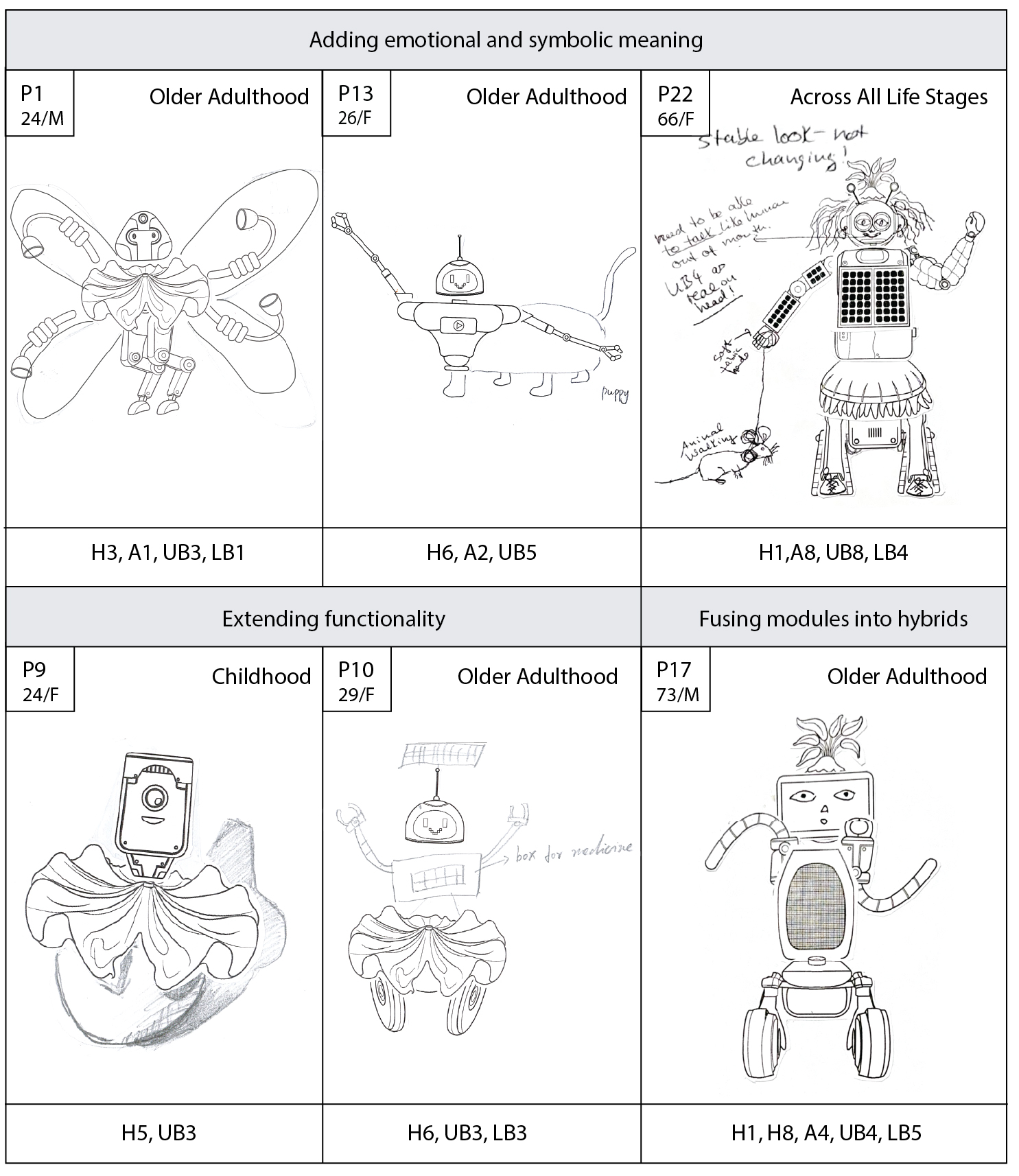}
    \caption{Customizations beyond provided components}\label{fig:figcustomization}
    \Description{A 2×3 grid of participant sketches showing robots that add symbolic elements (e.g., wings, pet-like forms), extend functionality, or fuse multiple modules together, demonstrating customizations that go beyond the original component set.}
\end{figure*}
Another strength of modular design was its ability to support customization that went beyond the parts we provided. Although participants were given a fixed set of modules in the co-design workshops, many extended the system with their own drawings or adaptations, treating modules not as finished products but as flexible design materials. In both workshops, 10 participants expanded the original set: some altered existing modules, while others invented entirely new ones. These creative extensions highlight how modular robots can spark imagination, allowing users to personalize robots in ways that reflect not only practical needs but also symbolic aspirations.

\textbf{Extending functionality.} Several participants added modules to meet specific functional needs at different life stages. For instance, in childhood designs, P12$_{\text{26/M}}$ added a ball-shaped lower body, explaining: \textit{“The ball (lower body) can move around.”} This improvised addition reflected children’s desire for play and dynamic mobility. In later life, functional expansion was tied more to care and protection. P10$_{\text{29/F}}$ emphasized long-term support, explaining: \textit{“A box to hold things like medicine, water, and food. I need it to help me carry items, like a shopping cart.”} She modified UB6 into a storage compartment and adapted the arms to handle supplies, envisioning a sustainable caregiver that could carry necessities and assist with errands. These designs show how participants personalized robots by embedding practical support for the distinct challenges of each life stage.

\textbf{Adding emotional and symbolic meaning.} Beyond functional extensions, participants customized robots to reflect values, emotions, and imagined futures. P1$_{\text{24/M}}$ drew large wings on his older adulthood design, inspired by his grandmother: \textit{“Mine was inspired by my grandma. She’s a very free spirit, and she enjoys moving around, but her legs make it difficult. So my focus was on mobility.”} The wings symbolized freedom and independence, reframing the robot as a metaphorical extension of dignity in aging. Similarly, P13$_{\text{26/F}}$ drew a puppy: \textit{“I draw a puppy because just in case in the future, I don't have kids, or grandchildren. I can have a pet like a dog. I can have a friend, like a robot friend.”} Her design shows how customization safeguarded imagined futures by ensuring companionship. This theme continued in P22$_{\text{66/F}}$’s design, where she sketched the robot walking her pet, incorporating animal care into its role. These examples illustrate that customization was not limited to practical assistance but also carried deep symbolic weight, embedding personal storytelling into design.

\textbf{Fusing modules into hybrids.} Some participants went further by combining existing modules into hybrid forms that produced new functions. P17$_{\text{73/M}}$ merged two heads (H1 and H8), layering a screen head with an expressive face. This fusion created a robot capable of both displaying information and conveying emotions, blending functional communication with affective presence. Such reconfigurations illustrate how participants reimagined modular parts as raw materials for innovation rather than fixed categories.

These examples demonstrate that modular design enabled personalization at multiple levels: extending functional capacity, embedding emotional symbolism, and hybridizing parts into new forms. By inviting users to treat modules as open-ended materials, modular robots became more than tools for immediate tasks. They emerged as living artifacts of personal imagination that could carry stories, reflect identities, and evolve with users’ envisioned futures.

\subsection{Adaptability through Modular Design}
While personalization highlighted how robots could be tailored differently across individuals, adaptability referred to their capacity to flexibly evolve with the same user over time. Participants consistently described adaptability as the quality that prevented a robot from becoming “frozen” in one role, instead enabling it to transform as circumstances shifted. As Participant 22$_{\text{65/F}}$ explained: \textit{“I think it's a good thing that is really adaptable. They can adapt to whatever you need.”}  

This framing positioned adaptability as the deeper contribution of modular design. By allowing parts to be added, removed, or reconfigured, modularity enabled robots to remain relevant both in the short term—handling overlapping needs within the same stage of life and in the long-term supporting users as they move from childhood to adulthood and into older age. In other words, adaptability is what makes modularity more than surface-level customization: it turns robots into dynamic systems capable of sustaining continuity, preserving attachment, and extending usability across contexts and over decades.  

We discuss adaptability in two dimensions. First, \textit{adaptability within the same life stage}, where robots reconfigure to address multiple coexisting demands such as play and protection in childhood, productivity and care in adulthood, or independence and safety in older adulthood (see \autoref{Adapt1}). Second, \textit{adaptability for lifespan evolution}, where robots transform incrementally or radically as a single person’s priorities shift across life stages (see \autoref{Adapt2}). These findings highlight adaptability as the temporal strength of modular design, positioning modular robots as companions that can continuously evolve alongside their users.

\subsubsection{Adaptability Within the Same Life Stage} \label{Adapt1}
\begin{figure*}[t]
    \centering
    \includegraphics[width=0.5\linewidth]{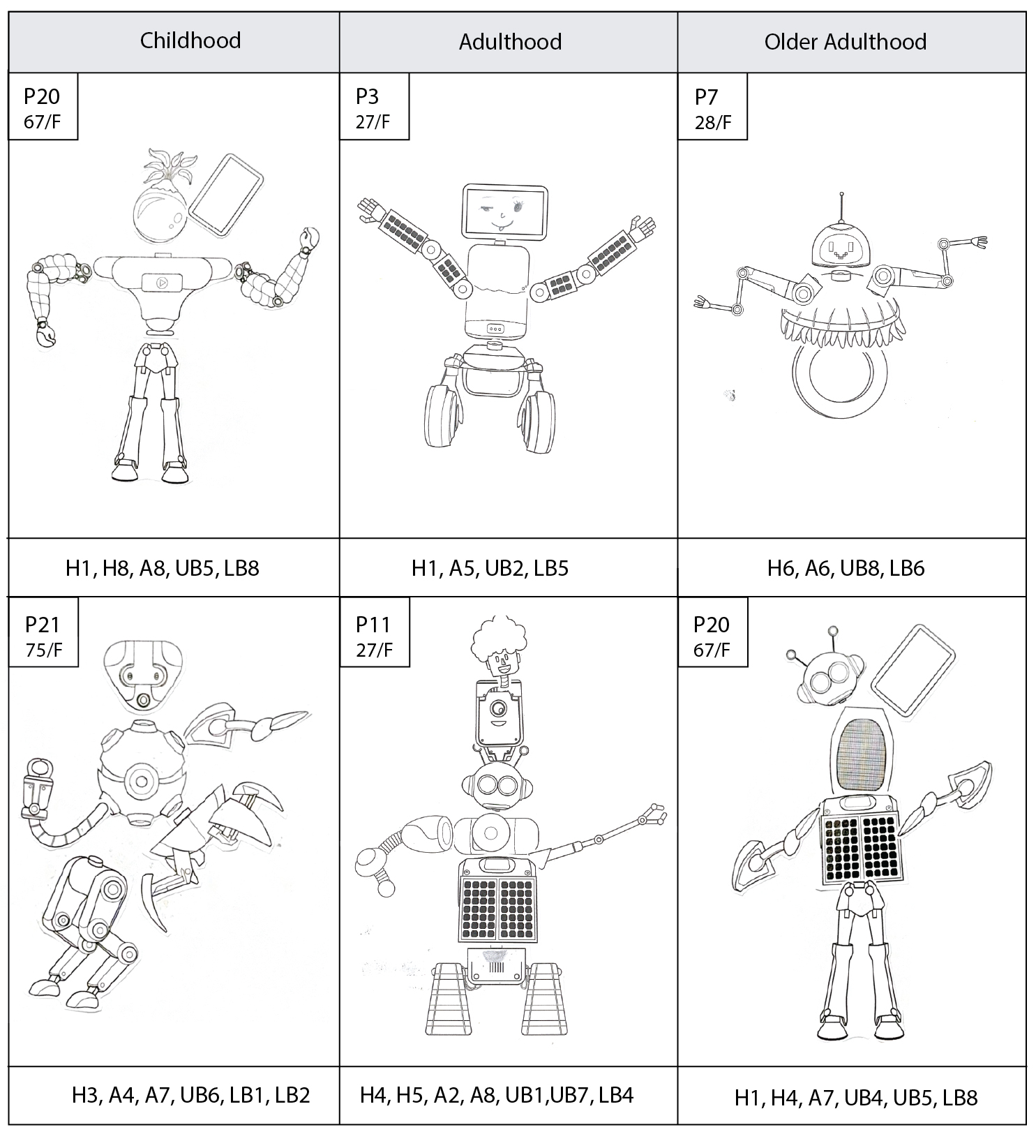}
    \caption{Adaptations within a single life stage}\label{fig:figlifestage}
    \Description{A 2×3 grid of participant sketches showing how participants adapted a single robot within one life stage (childhood, adulthood, or older adulthood) by changing modules to support different roles and activities.}
\end{figure*}
Unlike personalization, which highlighted differences across individuals, adaptability emphasized how the same person may require a robot to flexibly switch between roles in a single stage of life. Participants imagined everyday routines as complex and multi-layered, where supervision, play, work, care, and companionship might all be needed at once. Adaptability was therefore framed as the capacity to reconfigure modules fluidly in response to shifting tasks and contexts, ensuring one robot could keep pace with these overlapping demands.

\textbf{Childhood.} Participants described childhood as a period of curiosity, high energy, and vulnerability. Here, adaptability meant that a robot should simultaneously entertain, protect, and nurture a child, often within the same interaction. For instance, P20$_{\text{67/F}}$ combined modules for \textit{“supervision, safety, entertainment, hug”}, creating a system that could switch between watchful guardian, playful entertainer, and affectionate companion. P21$_{\text{75/F}}$ stressed that a child’s robot must \textit{“keep up with childhood energy,”} designing it with durable and mobile modules to adapt from active outdoor play to quieter educational support. These examples highlight that, in childhood, adaptability was about balancing protection and playfulness within the same system, which is also different from personalization, which would emphasize how one child’s robot differed from another’s.

\textbf{Adulthood.} In adulthood, adaptability was tied to competing responsibilities: work, home, and emotional life often collided in the same day. Robots were expected to flexibly shift between professional collaborator, domestic helper, and emotional supporter. P3$_{\text{27/F}}$ configured modules for information access, multitasking arms, and mobility, explaining that adulthood required a robot to \textit{“help me stay efficient but also lighten the load when life feels overwhelming.”} P11$_{\text{27/F}}$ layered projection, sensing, and multitasking features, admitting: \textit{“I become greedy in my adulthood,”} recognizing that adult life demanded efficiency, sustainability, and social awareness simultaneously. Here adaptability meant dynamic switching, helping in the office, supporting at home, offering comfort at night, within the same user’s life stage. By contrast, personalization would only highlight how different adults emphasized different priorities.

\textbf{Older adulthood.} In older adulthood, adaptability was imagined as the ability to flexibly combine medical support, companionship, and safety as conditions change. Rather than a single static “elder-care robot,” participants wanted a system that could shift between roles as vulnerabilities emerged or receded. P7$_{\text{28/F}}$ envisioned a robot that could adapt between home companion and hospital assistant: \textit{“suitable for hospital/home care… an emotional companion and butler in home.”} P20$_{\text{67/F}}$ designed modules to \textit{“help with hearing, talking for socialization, energy saving, watch for falling,”} combining medical vigilance with social engagement. These designs show adaptability as the ability to reconfigure for independence one day, companionship the next, and safety in emergencies—all within the same user. Unlike personalization, which would focus on whether different elders prefer mobility or comfort, adaptability here emphasized the same elder’s shifting needs across daily contexts.

Adaptability within life stages was consistently framed as the ability of one robot to manage overlapping and evolving needs for the same person, blending play and safety in childhood, balancing productivity and care in adulthood, and supporting independence and protection in older adulthood. This stands in contrast to personalization: while personalization emphasized diversity across individuals, adaptability emphasized flexibility within individuals, making modular robots' dynamic systems that could fluidly adjust to the multifaceted realities of everyday life.

\subsubsection{Adaptability for Lifespan Evolution} \label{Adapt2}
\begin{figure*}[t]
    \centering
    \includegraphics[width=0.5\linewidth]{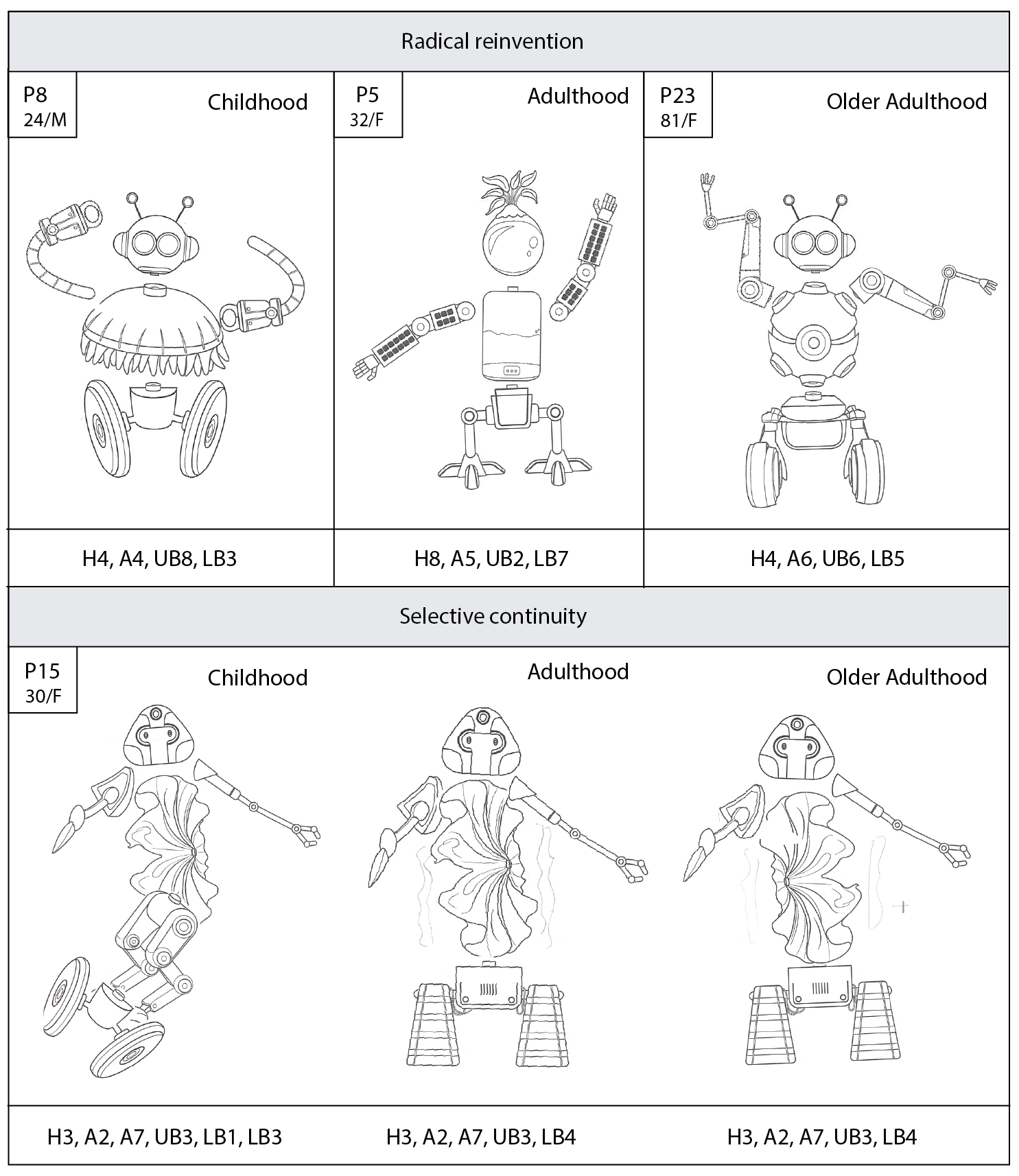}
    \caption{Adaptability over life stages}\label{fig:figadapt}
    \Description{A 2×3 grid of participant sketches showing two adaptability patterns over the lifespan: radical reinvention of form between childhood, adulthood, and older adulthood, and selective continuity where some modules remain stable while others change.}
\end{figure*}
Adaptability for lifespan evolution emphasized how the same robot could change alongside one individual over decades. Participants valued the possibility of selectively upgrading parts while retaining familiar modules, so the robot could grow with them without losing its recognizable identity. Adaptability here was imagined in two main ways: radical reinvention, wherein the robot’s form and function shifted entirely at each stage, and selective continuity, where certain modules remained constant while others were swapped to reflect new needs. Both modes highlight modularity as a mechanism for continuity across the lifespan, turning robots into long-term companions rather than disposable tools.

\textbf{Radical reinvention.} Many participants imagined their robots transforming completely from one stage to the next. P8$_{\text{24/M}}$ began with playful, expressive modules in childhood (H4, A4, UB8, LB3), but in adulthood replaced them with plant-like and storage modules (H8, A5, UB2, LB7) to emphasize functionality and practicality. By older adulthood, his design shifted again to stability and support, with H4, A6, UB6, and LB5. His robot thus moved from whimsical playmate, to grounded assistant, to mobility partner, demonstrating adaptability as radical redesign to match shifting life priorities. Similarly, P6$_{\text{25/F}}$ described her robot as a playful entertainer in childhood (\textit{“replicate the abilities of a child”}), a multitasking professional tool in adulthood (\textit{“let me have two tasks at one move”}), and a practical helper in older adulthood (\textit{“helpful”}). These examples show how adaptability allowed participants to imagine one robot reinventing itself completely to embody different roles across life stages.

\textbf{Selective continuity.} Other participants preferred to preserve core modules across stages, using continuity to sustain familiarity and attachment while selectively adapting functions. P15$_{\text{30/F}}$ illustrates this trajectory. In childhood, she emphasized protection and activity with H3, A2, A7, UB3, and LB4. In adulthood, she retained UB3 and LB4 but reframed them for professional contexts, adapting her robot from playful protector to efficient workplace assistant. In older adulthood, she again kept H3, A7, UB3, and LB4, but added comfort features (\textit{“a cushion to massage my lower back, extra support on my joints, and a headset to help me sleep”}). Here the same modules gained new meanings across contexts, showing how adaptability allowed continuity of identity while still accommodating shifting needs. P5$_{\text{32/F}}$ offered another example: her childhood teddy-bear robot with soft, plush arms evolved into an adulthood companion for stress relief, retaining the motif of softness but reinterpreting it as relaxation rather than play. Such designs illustrate adaptability as a process of reinterpretation, wherein familiar elements anchor emotional connection while new modules address emerging demands.

Adaptability for lifespan evolution thus captured a temporal quality distinct from personalization, whereas personalization highlighted how different people customized robots in unique ways, adaptability described how one person could sustain a relationship with a single evolving robot across decades. Whether through radical reinvention or selective continuity, participants envisioned modular robots as companions that grow with them, preserving familiarity while transforming to meet changing priorities. This capacity for continuity and reinvention highlights adaptability’s significance: it not only supports long-term usability and ecological sustainability, but also deepens emotional attachment by enabling robots to become lifelong partners rather than disposable machines.

\subsection{Sustainability through Modular Design}

Sustainability in our study was not only about reducing waste, but about ensuring that robots remain relevant, meaningful, and ecologically responsible over the long term. Participants consistently described modularity as the mechanism that could prevent obsolescence, keeping a robot in use by allowing broken or outdated parts to be swapped, reused, or reinterpreted. Unlike personalization (focused on individual differences) or adaptability (focused on shifting demands), sustainability highlighted how modular design supports both the material longevity of robots and the emotional continuity of human–robot relationships.  

We found that participants articulated sustainability in two intertwined ways. First, they emphasized the \textit{material dimension}, seeing modularity as a way to repair, reuse, and evolve robots without discarding the whole system (see \autoref{sus1}). Second, they underscored the \textit{emotional dimension}, describing how continuity, attachment, and identity could sustain long-term human–robot relationships (\autoref{sus2}). These two dimensions positioned sustainability as the deepest temporal promise of modular design, ensuring that robots are not short-lived devices but enduring companions that grow with their users over decades.

\subsubsection{Material Sustainability: Repair, Reuse, and Evolve} \label{sus1}

Material sustainability was consistently described as the ability to repair, reuse, and evolve robots without discarding the entire system. Participants emphasized that current consumer technologies often force users into full replacement even when only one part fails, generating both unnecessary cost and waste. In contrast, they framed modularity as a way to break this cycle by normalizing repair and reuse, everyday practices rather than rare exceptions. As P22$_{\text{66/F}}$ criticized, \textit{“Robots on the market often require buying separate devices for different functions, which would be very expensive. Having parts that can be exchanged, and swapped out when they no longer fit your needs is a valuable functionality.”} Her comment underscores frustration with today’s “single-use” logic and highlights how modular robots could support more sustainable consumption practices.

\textbf{Repair.} Participants often positioned repair as the first line of defense against waste. They imagined scenarios where a broken module—whether a mobility base, sensor, or arm—could simply be replaced while the rest of the robot remained intact. As P20$_{\text{67/F}}$ explained, \textit{“Having parts to exchange is always better than throwing the whole thing away. That’s how our technology works right now. If something breaks, you basically just throw it away.”} Her observation reflects broader concerns with contemporary consumer electronics, where repair is rarely feasible. By contrast, modular robots offered a vision wherein repair was built into the system from the start, ensuring longevity by design.

\textbf{Reuse.} Beyond repair, participants also saw opportunities for modules to be creatively repurposed once they no longer served their original purpose. This kind of reuse extended the value of components across multiple lifecycles. For instance, P10$_{\text{29/F}}$ suggested that \textit{“an old mobility base can be reused as a storage function.”} Instead of disposal, outdated parts could find new roles in everyday life. This approach reframed sustainability not only as resource conservation but as imaginative reconfiguration, where older modules still carried value as building blocks for new functionality.

\textbf{Evolution.} Finally, participants highlighted evolution as the most significant material contribution of modularity. Rather than seeing robots as disposable devices, they imagined them as systems that could continuously upgrade and adapt across a person’s lifetime. P2$_{\text{21/M}}$ described this vision clearly: \textit{“I would prefer a modular robot which can be upgraded across a person's lifetime so it builds a connection with robot instead of replacing it.”} Evolution here referred not just to technical updates, but to the possibility of maintaining continuity and history with the same robot. Each replacement part became part of a longer trajectory, allowing the robot to grow with its user instead of being abandoned.

These perspectives show how modular design transforms material sustainability from an afterthought into a central principle of robotic systems. Repair prevents waste in the short term, reuse gives old parts new lives, and evolution sustains the system across decades. By embedding these practices into the core of design, modular robots directly address the problem of technological obsolescence, ensuring that robots remain materially viable and ecologically responsible over the long run.

\subsubsection{Emotional Sustainability: Continuity, Attachment, and Identity} \label{sus2}

Emotional sustainability was just as critical as material sustainability. Participants emphasized that keeping familiar modules over time gave robots a sense of continuity, turning them into long-term companions rather than disposable gadgets. This continuity allowed robots to preserve familiarity and accumulate shared meaning as they accompanied people through different stages of life. 

\textbf{Continuity.} Participants described continuity as the ability to maintain recognizable elements of a robot over time, ensuring that it felt stable even as other parts changed. As P7$_{\text{28/F}}$ reflected: \textit{“It kind of becomes part of your everyday life, and then you share memories just by looking at it, because you still have some old pieces or old modules that you will recognize from before—nostalgia.”} For her, sustainability meant more than functionality: it meant the ability to look at an old module and remember past experiences. P16$_{\text{73/F}}$ similarly described repair as a way of maintaining expectations and relationships, explaining: \textit{“I would like to repair a current robot and keep continuity and expectations. It would grow old with me, like a friend.”} In both cases, continuity was not about avoiding change altogether but about retaining enough familiarity to preserve the thread of a shared history.  

\textbf{Attachment.} Continuity often deepened into attachment. Participants compared modular robots to cherished toys or family members that carry memories across time. P13$_{\text{26/F}}$ illustrated this connection by saying, \textit{“Sometimes I still keep the toy. So if you change the robot, it’s just changed part of it, like keeping a family member.”} By preserving recognizable modules, participants imagined robots as repositories of personal and familial memories, where even a single arm or torso could serve as a reminder of earlier life stages. Attachment also meant participants were more willing to maintain and repair robots rather than replace them, echoing sustainability research showing that emotional bonds extend product lifetimes. This attachment anchored robots as durable companions rather than short-lived devices.  

\textbf{Identity.} For some, sustainability was expressed most strongly as identity continuity. Participants described modular robots not only as companions but as extensions of themselves, whose meaning grew as their own lives unfolded. P15$_{\text{30/F}}$ explained: \textit{“It felt like it's an extension of me, but not something external that I'm acquiring. So having the ability to move the pieces would feel like I'm still onto the same person or like the same me, but it's just evolving as the time is changing as opposed to like just buying a new phone every time and I'm not attached.”} This “extension of self” framing highlights sustainability as more than ecological efficiency: it is about ensuring that robots remain part of a person’s identity across time. P1$_{\text{24/M}}$ echoed this sentiment by likening his robot to Spider-Man’s suit, describing it as an “attachment” that became part of who he was. In these accounts, sustainability was achieved not simply by preserving materials but by sustaining identity, trust, and recognition across life stages.  

These accounts show sustainability as the deepest promise of modular robots. \textit{Materially}, modularity reduces waste by making repair, reuse, and evolution standard practices. \textit{Emotionally}, it preserves continuity, strengthens attachment, and embeds robots into users’ sense of identity. By combining these two dimensions, modular robots gain the capacity to endure not only as recyclable systems but as long-term companions. Sustainability thus emerges as a central design goal, ensuring that robots remain relevant, cherished, and ecologically responsible across the human lifespan.

\section{Discussion}
Our study contributes to HRI and sustainable design research by showing how modularity enables three interdependent qualities: personalization, adaptability, and sustainability, which together shape long-term human–robot relationships. Prior research has often treated these qualities in isolation: personalization as short-term customization of robot behaviors \cite{Bahar2019Personalization}, adaptability as technical self-reconfiguration without user input \cite{yim2000polybot}, and sustainability as ecological efficiency detached from lived experiences \cite{blevis2007sustainable, scheutz2021envirobots}. By contrast, our findings demonstrate how these dimensions reinforce one another when users themselves act as co-designers. Personalization provides the emotional entry point, adaptability ensures continuity through changing needs, and sustainability emerges from the interplay of material repair and emotional attachment. These insights reposition modularity not only as an engineering feature but as a socio-technical mechanism that supports meaningful, lifespan-oriented companionship.  

In what follows, we first articulate how these three dimensions function as mutually supporting processes (see \autoref{dis1}), showing modularity’s integrative role as a design philosophy. We then translate these insights into actionable design principles (\autoref{dis2}) that can guide future efforts in building robots that are ecologically responsible, emotionally resonant, and enduring across the human lifespan.

\subsection{Mutual Support through Modular Design: Personalization, Adaptability and Sustainability} \label{dis1}

Our findings indicate that personalization, adaptability, and sustainability are not isolated qualities but mutually reinforcing dimensions of modular robot design. This integration addresses a persistent gap in prior HRI and design research. Much of the existing work has examined these dimensions in isolation: personalization has been treated as a question of user preference or short-term customization \cite{Bahar2019Personalization, lacroix2023s, lee2012personalization}, adaptability has oftentimes been reduced to technical self-reconfiguration without user involvement \cite{yim2000polybot}, and sustainability has oftentimes been framed in terms of energy efficiency or material recycling rather than long-term relational continuity \cite{blevis2007sustainable, scheutz2021envirobots}.

The only design-theoretical attempt to integrate these strands so far is the recent Sustainability, Adaptability, and Modularity (SAM) framework \cite{chen2025sustainable}. SAM usefully highlights the interdependence of these three qualities, positioning sustainability as the overarching aim, modularity as the enabling mechanism, and adaptability as the dynamic capacity to evolve with users’ needs. Our work takes this agenda as a starting point and asks how it plays out when people actually co-design modular robots for their own lives.

Our contribution lies in complementing SAM along three dimensions. First, we recognize SAM as a design-theoretical framework that explicitly interlinks sustainability, adaptability, and modularity, and we reaffirm its value for lifespan-oriented HRI. At the same time, our analysis highlights a specific limitation: SAM articulates sustainability and adaptability as desirable system properties, but does not make clear for whom or how these properties are oriented toward particular life courses. Second, we address this limitation by introducing \textit{personalization} as a human-centered process that orients SAM’s system-level aims toward specific people and life situations. In our data, modular configurations were consistently tied to participants’ concrete roles, relationships, bodies, and constraints. We use personalization to describe how a robot’s changing form and behavior are continually adjusted to these situated values and priorities. Our findings suggest that, in lifespan-oriented HRI, the connection between adaptability and sustainability is strongly mediated by meaning-making---how people come to see a robot as “for me,” “part of my life,” and “worth keeping.” When modular reconfigurations express identity, roles, and values, they cultivate ownership and attachment, which in turn increases people’s willingness to maintain, repair, and carry modules forward rather than abandon the robot. We therefore treat personalization not as an add-on dimension, but as an important human-centered mechanism that gives direction to adaptability and motivates sustained use. Without this lens, adaptability remains directionless and sustainability risks being understood only in technical terms (e.g., longer lifetimes) rather than as something that is worth maintaining and repairing. In PAS, modularity therefore functions not only as a technical enabler but as a socio-technical bridge that connects personalization, adaptability, and sustainability when people are positioned as co-designers. Third, we provide the first empirical grounding of SAM through modular co-design workshops, demonstrating how people themselves assemble, reconfigure, and sustain robots in ways that actualize the framework’s principles. These contributions move SAM from a conceptual vocabulary to an actionable design orientation, and distinguish our PAS framework as a human-centered, empirically grounded account of how lifespan-oriented robot design can be realized in practice.

\textbf{Personalization as the foundation.} We found that personalization emerged as the entry point for modular engagement. By choosing, skipping, recombining, or inventing modules, participants created robots that embodied their own values, preferences, and priorities. This goes beyond prior HRI research that has focused on tailoring robot roles for specific demographics (e.g., children, older adults) \cite{klamer2010acceptance, kory2019exploring}. Instead, personalization in our study highlighted the agency of individuals to define what a robot should be, even within the same demographic group. Such active shaping established a sense of ownership and attachment, laying the emotional groundwork for later practices of repair rather than disposal.  

\textbf{Adaptability as dynamic extension.} Once this initial sense of belonging was established, adaptability provided the mechanism for continuity. As people’s needs shifted, whether within one stage of life or across decades, robots could be flexibly reconfigured through modularity. Unlike prior design research that tends to focus on a robot’s role at one moment in time (e.g., play in childhood, care in old age) \cite{broadbent2009acceptance, tanaka2007socialization}, our participants envisioned the same robot accompanying them through multiple transitions. Adaptability thus emerged not only as a technical capacity, but as an human-centered design strategy: allowing the robot to grow and evolve with its owner. This extends prior HCI work on personalization into a temporal domain, showing how co-design processes can surface expectations for long-term, rather than episodic, adaptation.  

\textbf{Sustainability as the outcome.} Sustainability was not introduced to participants as an explicit design goal. Instead, it surfaced in how they described keeping, repairing, and repurposing a single robot over time. Materially, we understand that modularity enables repair, reuse, and incremental upgrading, reducing e-waste and extending lifecycles. Emotionally, the participants framed continuity, memories, and attachment as reasons to preserve a familiar robot instead of replacing it with a newer model. In their reflections, participants often described these decisions in terms of making one robot last, keeping a trusted companion, and avoiding unnecessary replacement. We therefore interpret these practices as a form of sustainability, reading them through a sustainable HCI lens. Unlike sustainability research in HRI that has often emphasized ecological tasks performed by robots (e.g., prompting people to recycle) \cite{scheutz2021envirobots}, our findings highlight how the design of the robot itself can embody sustainable principles by fostering both ecological responsibility and emotional continuity. This perspective resonates with work in HRI that examines robots as mediators of destruction, catharsis, and emotional release rather than purely utilitarian tools \cite{luria2020destruction}. This suggests that long-lived companions must be able to accommodate intense, sometimes messy interactions without being treated as disposable. This demonstrates that sustainable HCI requires integrating material and emotional dimensions rather than treating them separately.  

\textbf{Interrelation and modular design’s role.} These three dimensions formed a reinforcing cycle. Personalization initiated attachment, which made adaptability meaningful; adaptability ensured continuity, which in turn deepened attachment and prevented replacement; and both together culminated in sustainability, where ecological responsibility was inseparable from emotional endurance. Modularity was the key mechanism enabling this cycle. By providing interchangeable building blocks, it supported personalization at the outset, adaptability across changing contexts, and sustainability over the long term. In this sense, modular design does not simply enable technical flexibility: it underpins a relational ecology in which robots remain relevant, meaningful, and durable companions across the human lifespan.  

This perspective not only fills a gap in current HRI design research, but also provides a framework for rethinking how robots can be designed as enduring companions rather than short-lived devices. By treating personalization, adaptability, and sustainability as mutually supporting, we foreground modularity as both a technical mechanism and a design philosophy that binds ecological responsibility with human attachment in long-term human–robot relationships.

\subsection{Design Principles for Future Sustainable HCI/HRI} \label{dis2}

Our findings suggest that modularity can serve as a design philosophy for creating robots that remain materially viable and emotionally meaningful over decades. To translate these insights into practice, we outline several design principles that extend current approaches in HCI and HRI toward sustainable, lifespan-oriented interaction.

\textbf{Support Inclusive Personalization Across Life Stages.}  
Designers and researchers should treat personalization as a foundation rather than a superficial feature. Our study showed that participants consistently wanted robots to reflect their own priorities, values, and circumstances, whether as children, adults, or older adults. For children, this means prioritizing features that stimulate curiosity, outdoor play, and independent learning rather than encouraging passivity or over-assistance. For adults, modules should be able to flexibly support the coexistence of multiple roles from workplace productivity to household assistance and emotional support. For older adults, designs should emphasize mobility, safety, and companionship, ensuring independence, dignity, and emotional well-being. Personalization in modular robots thus enables robots to feel less like standardized products and more like extensions of individual life paths.  

\textbf{Provide Open Platforms for Co-Creation.}  
Designers and researchers should move beyond closed, one-size-fits-all systems and instead create open modular platforms that invite users to actively participate in construction and evolution. In our workshops, participants extended the provided modules with new sketches, hybrid designs, and symbolic features, treating modules as open-ended design materials rather than fixed parts (Section 5.1.2). This finding highlights the need for platforms that empower users to experiment, remix, and evolve robots over time. Such openness not only fosters agency and stronger identification with robots but also situates users as co-creators rather than passive consumers, encouraging deeper and more enduring engagement \cite{desmet2007framework}, \cite{alves2021collection, cui2024robotic}.  

\textbf{Design for Adaptability Within and Across Life Stages.}  
Adaptability should be treated as more than technical flexibility; it is the lived capacity of robots to evolve with the same user’s shifting demands in both function and form. Our findings showed that users often required different roles within a single stage of life, balancing play and safety in childhood, productivity and care in adulthood, or independence and companionship in older adulthood. At the same time, adaptability should guard against robots that look and behave exactly the same for decades. Companions that never change can begin to feel oddly out of sync with people’s lives, raising a sense of “social uncanniness” when they do not age or transform alongside their partners \cite{hoffman2020social}. Designing modules that can be added, removed, or visually refreshed over time allows robots to register life transitions—subtle shifts in appearance, capabilities, or interaction style that mirror peoples’ own changing identities. Designers should therefore create modules that allow dynamic switching between roles within one life stage, while also enabling smooth transitions across stages. This principle ensures that robots are not frozen in one role but can accompany people through multiple transitions, reflecting the realities of complex and evolving lives.  

\textbf{Foster Sustainability as Both Material and Emotional Practice.}  
Sustainability in modular robots requires addressing both material longevity and emotional continuity. Materially, designs should facilitate easy repair, repurposing, and upgrading of individual modules to prevent full-system replacement and reduce e-waste. Emotionally, continuity can be supported by allowing certain modules or forms to persist across upgrades, enabling users to carry memories and identity connections into the future. As participants noted, familiarity with old modules can evoke nostalgia and strengthen attachment, making robots feel like companions rather than disposable tools. Evidence from care settings with the therapeutic robot PARO similarly shows that long-term engagement with a single, familiar robot can help reduce agitation and loneliness and support more positive care experiences for older adults \cite{6650427, hung2019benefits}. Designers and researchers should therefore enhance the perceived value of robots by encouraging practices of repair and retention, creating products that grow more meaningful over time rather than depreciating with obsolescence \cite{chapman2009design, hebrok2014design}.  

\textbf{Embrace Modularity as a Mechanism for Lifespan-Centered Design.}  
Ultimately, the philosophical value of modularity extends beyond technical efficiency. Modularity enables users to imbue robots with personal meaning, ensures their relevance across multiple life stages, and supports ecological responsibility through continuity of use. By integrating personalization, adaptability, and sustainability into a mutually reinforcing cycle, modular design allows robots to evolve into long-term companions, that are extensions of the self that carry both practical utility and symbolic significance. This principle challenges designers to think not only about immediate usability but about how robots can endure as artifacts of shared histories, identity, and care, sustaining human–robot relationships across decades.  

\subsection{Limitations and Future Directions}

This study concentrated on younger and older adults, as adulthood represents a continuous period marked by significant functional, social, and emotional transitions \cite{carstensen1992social, english2014selective}. This focus allowed us to explore how modular robots could adapt to evolving expectations of autonomy, caregiving, health maintenance, career transitions, and retirement. While this adult-centered lens enabled us to articulate practical design guidelines for adaptability and sustainability, it also introduced limits. Younger adults generally reported no difficulty envisioning robots for older life stages, but several elderly participants struggled to imagine robots in their own youth, citing mismatches with the domestic and occupational environments of that era. These differences highlight both the complexity of cross-generational design and the need for broader, more inclusive perspectives in future research.

Although childhood was included as a design scenario, children themselves did not participate in our workshops, since we believe that understanding adult and older adult perspectives first can better inform a future study that includes children as participants. All childhood designs were therefore based on adult projections rather than children’s lived experiences. This constrains the ecological validity of our findings for early life stages, especially around play, learning, and safety. We began with this adult-centred, lifespan sample as an experimental first step in developing the PAS framework: adults can speak both as former children and as current or prospective caregivers, which allowed us to articulate and stress-test the dimensions of personalization, adaptability, and sustainability within a single, coherent lifespan narrative. At the same time, this vantage point means that our accounts of early life stages should be read as provisional and situated in adult imaginaries, rather than as direct evidence of children’s perspectives. Future work should complement this view by directly involving children and teenagers—for example, through short, age-appropriate co-design activities or probes—to triangulate and extend our findings about how modular robots might support early-life play, learning, and safety.

Our workshops primarily centered on the functional roles of modular robots, yet several participants expressed interest in aesthetics and embodiment, such as human-like or culturally resonant appearances. This points to a broader gap in HCI and HRI---while much work has addressed system-level interactions, comparatively little attention has been paid to the design of individual components and their integration into coherent, meaningful forms \cite{disalvo2002all}. At the same time, our contribution is grounded in conceptual and graphical co-design probes rather than a tangible modular prototype. Participants imagined how a single robot might adapt with them over decades, but neither embodiment nor technical feasibility was tested. As a result, our claims about lifelong adaptability and modular growth should be understood as design-oriented and hypothetical rather than as evidence of a working system. To substantiate these arguments, future research should include material explorations and iterative prototyping—even at low fidelity—that examine how specific module designs, connectors, textures, and forms perform in real use, and how they align with peoples’ aesthetic values, cultural expectations, and emotional identities. Such work would help clarify which aspects of personalization, adaptability, and sustainability are practically achievable, and what trade-offs arise when modular robots are implemented in concrete contexts.

Moving forward, future research should also reconsider the structural assumptions built into the modules themselves, and further relax or even abandon the humanoid scaffold that underpins our current module set. In this study, we deliberately started from a body-like construction grammar precisely to surface how participants reproduce or resist familiar expectations about “humanoid” robots. Our findings show that people are willing to bend and break this structure when modularity allows them to embed additional functions and meanings. Building on metaphor-driven approaches that use open-ended prompts to expand robot imaginaries beyond human-like or pet-like archetypes, such as Alves-Oliveira et al.’s collection of metaphors for HRI \cite{alves2021collection}, future co-design studies could instead begin from non-anthropomorphic, environment-centered, or distributed module grammars. This would allow us to investigate how lifespan-oriented modularity plays out when the underlying structure itself is no longer tied to a human body form.

\section{Conclusion}

This paper introduced two co-design workshops that explored how modular robots could accompany people across their lifespans, from childhood to adulthood and into older age. By engaging participants in co-design, we identified three interdependent qualities that modularity affords in human–robot interaction: personalization, adaptability, and sustainability. Personalization enabled participants to construct robots that reflected their own values, needs, and imaginations, establishing ownership and emotional resonance. Adaptability extended this foundation by allowing robots to flexibly transform within and across life stages, responding to shifting priorities without losing continuity. Sustainability emerged at the intersection of these two capacities: materially, by enabling repair, reuse, and evolution of components, and emotionally, by fostering attachment, continuity, and identity across decades.  

Together, these findings show that modular design is not simply a technical architecture but a human-centered mechanism for building long-term, meaningful, and ecologically responsible relationships with robots. Unlike traditional robots that often become obsolete or are discarded, modular robots can evolve with their users, transforming from playful companions into supportive caregivers while retaining familiar elements that preserve continuity. This repositions robots not as short-lived tools but as enduring companions with the potential to achieve what participants described as “extension of self” or even “heirloom” \cite{blevis12007ensoulment, jung2011deep, odom2009understanding} status.  

For HCI and HRI, our study highlights the importance of designing robotic systems that are open, flexible, and co-created with users. By foregrounding personalization, adaptability, and sustainability as mutually reinforcing dimensions, modular design provides a pathway for creating robots that are inclusive, resilient, and deeply integrated into the rhythms of human life. Future research and design practice should continue to develop modular platforms that empower users to shape, reconfigure, and sustain their robots over time, ensuring that these systems remain both technologically relevant and emotionally meaningful across the human lifespan.

\bibliographystyle{ACM-Reference-Format}
\bibliography{references}

\appendix
\vspace{9cm}
\section{Full Lifespan Co-Design Outcomes}
\label{appendix:all_sketches}
\begin{figure*}[h]
    \centering
    \includegraphics[width=\linewidth]{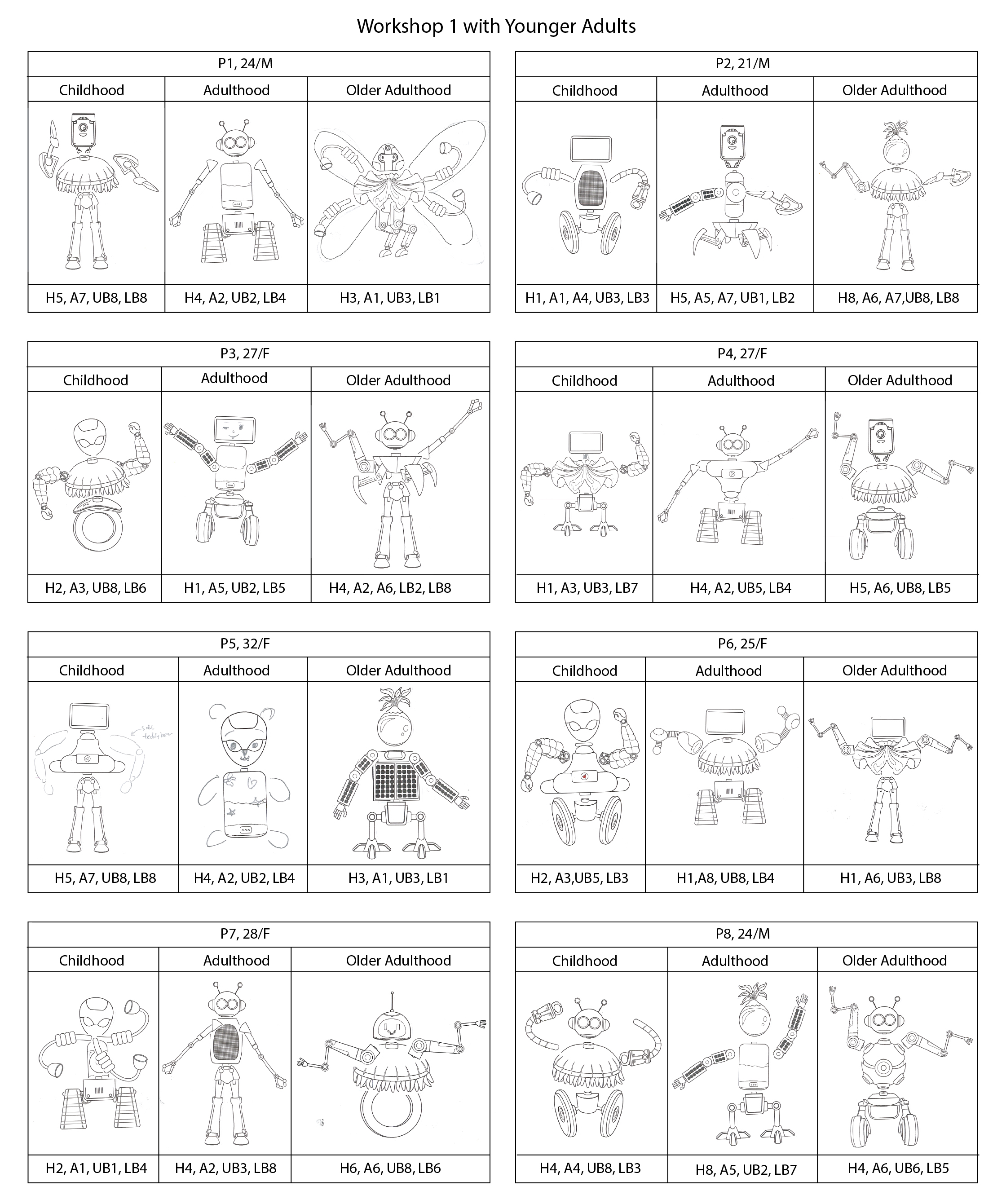}
    \caption{P1–P8's lifespan modular robot design outcomes from the younger adults workshop}
    \Description{This figure presents the full set of lifespan modular robot designs from younger adult participants P1–P8, showing each person’s childhood, adulthood, and older adulthood robot configurations side by side.}
    \label{fig:drawing01}
\end{figure*}

\begin{figure*}[h]
    \centering
    \includegraphics[width=\linewidth]{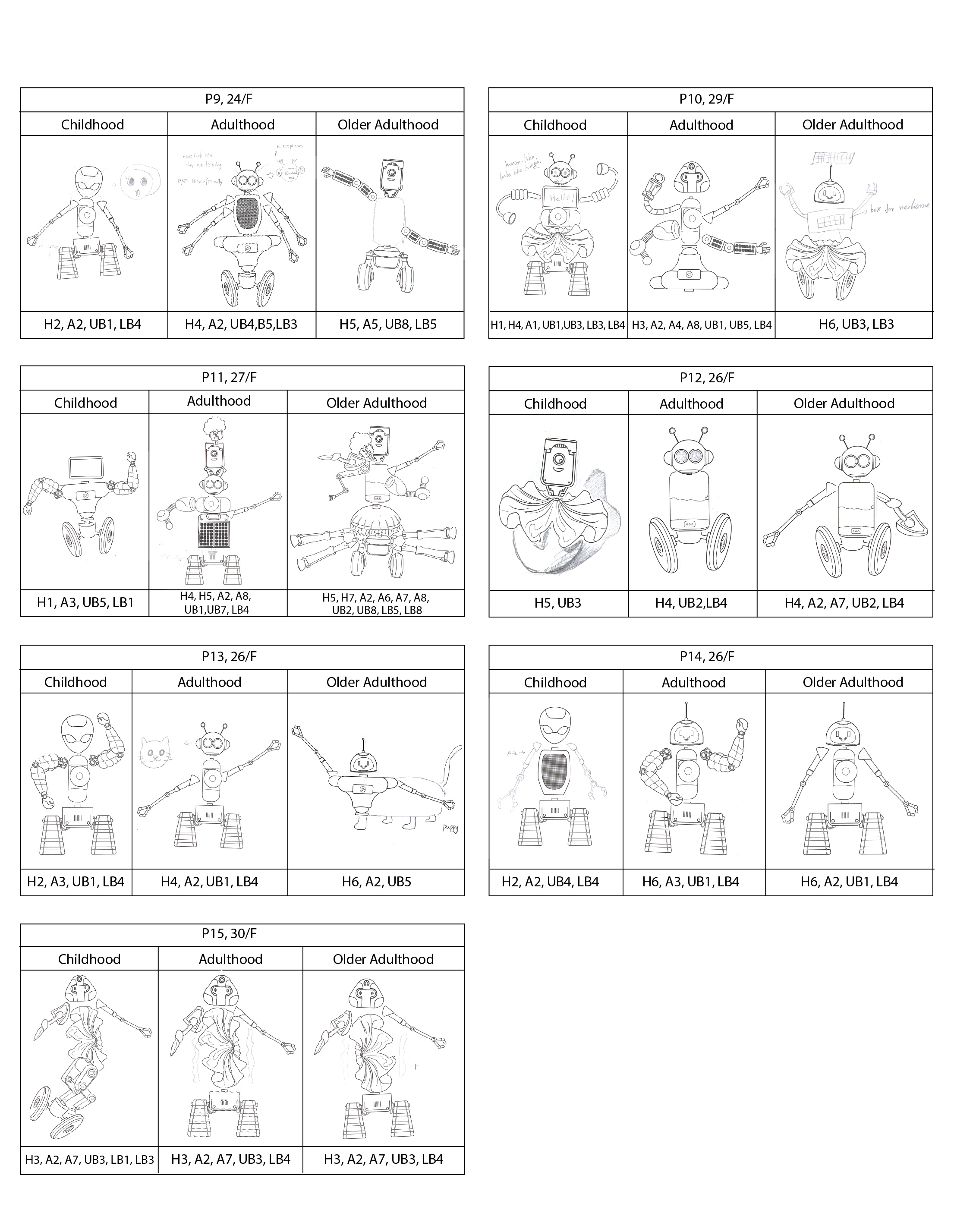}
    \caption{P9–P15's lifespan modular robot design from the younger adults workshop}
    \Description{This figure presents the remaining lifespan modular robot designs from younger adult participants P9–P15, again displaying each participant’s childhood, adulthood, and older adulthood robot configurations in a grid.}
    \label{fig:drawing02}
\end{figure*}

\begin{figure*}[h]
    \centering
    \includegraphics[width=\linewidth]{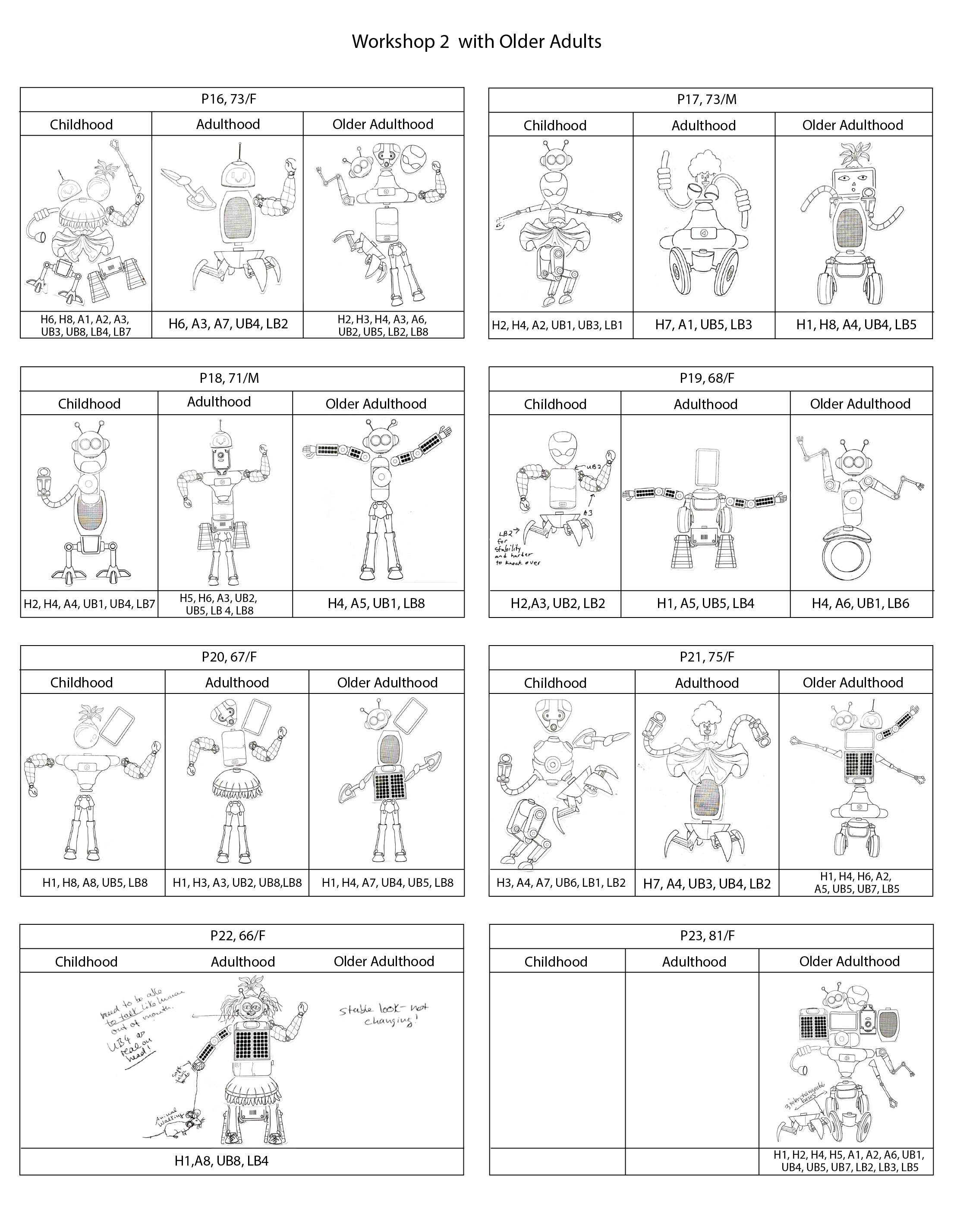}
    \caption{P16–P23's lifespan modular robot design from the older adults workshop}
    \Description{This figure presents the lifespan modular robot designs from older adult participants P16–P23, showing how each participant envisioned their robot across childhood, adulthood, and older adulthood.}
    \label{fig:drawing03}
\end{figure*}

\end{document}